\documentclass[aps,prd,twocolumn,superscriptaddress,groupedaddress,nofootinbib]{revtex4}
\usepackage{graphicx}  
\usepackage{dcolumn}   
\usepackage{bm}        
\usepackage{amssymb}   
\usepackage{amsmath}
\usepackage{xcolor}
\usepackage{bbold}
\usepackage{amsthm}
\usepackage{enumitem}
\usepackage{physics}
\usepackage{braket}
\usepackage{mathtools}
\usepackage{slashed}
\usepackage{marginnote}
\usepackage{stackengine}

\stackMath
\usepackage{tikz}
\usetikzlibrary{patterns}
\usepackage[compat=1.1.0]{tikz-feynman}

\usepackage[colorlinks]{hyperref}
\hypersetup{
     breaklinks=true,
    pdfstartview={FitH},    
    colorlinks=true,       
    linkcolor=blue,          
    citecolor=red,        
    filecolor=magenta,      
    urlcolor=blue,           
    anchorcolor=green,      
    linktocpage=true
}

\newcommand{\be}{\begin{equation}}\newcommand{\ee}{\end{equation}}
\newcommand{\bea}{\begin{eqnarray}}\newcommand{\eea}{\end{eqnarray}}
\newcommand{\brr}{\begin{array}}\newcommand{\err}{\end{array}}
\newcommand{\bit}{\begin{itemize}}\newcommand{\eit}{\end{itemize}}
\newcommand{\ben}{\begin{enumerate}}\newcommand{\een}{\end{enumerate}}

\newcommand{\ba}{\begin{array}}
\newcommand{\ea}{\end{array}}

\definecolor{darkred}{rgb}{.8,0,0}

\definecolor{darkblue}{rgb}{0,0,.7}

\def\1{{_{1}}}\def\2{{_{2}}}

\def\noHe0{:\;\!\!\;\!\!:H_e(0):\;\!\!\;\!\!:}
\def\noHm0{:\;\!\!\;\!\!:H_\mu(0):\;\!\!\;\!\!:}

\begin{document}

\title{Running of the Number of Degrees of Freedom in Quantum Conformal Gravity}

\author{Stefano Giaccari }
\email{stefano.giaccari@pd.infn.it}
\affiliation{Dipartimento di Fisica e Astronomia Galileo Galilei e INFN sez. di
Padova, Universit\`a di Padova, Via Marzolo 8, Padova, 35131, Italy}

\author{Petr Jizba}
\email{p.jizba@fjfi.cvut.cz}
\affiliation{FNSPE,
Czech Technical University in Prague, B\v{r}ehov\'{a} 7, 115 19, Prague, Czech Republic}

\author{Jaroslav K\v{n}ap}
\email{knapjaro@fjfi.cvut.cz}
\affiliation{FNSPE,
Czech Technical University in Prague, B\v{r}ehov\'{a} 7, 115 19, Prague, Czech Republic}

\author{Les{\l}aw Rachwa{\l} }
\email{L.Rachwal@fjfi.cvut.cz}
\affiliation{FNSPE,
Czech Technical University in Prague, B\v{r}ehov\'{a} 7, 115 19, Prague, Czech Republic}
\affiliation{Departamento de Física - Instituto de Ciências Exatas,
Universidade Federal de Juiz de Fora, 33036-900, Juiz de Fora, MG, Brazil}

\date{\today}

\begin{abstract}

We study how the number of degrees of freedom in Weyl conformal gravity runs with the energy scale from the UV fixed point. To this end we employ two approaches.  First, we utilize the Fradkin--Tseytlin prescription for the number of degrees of freedom and demonstrate that the one-loop result is highly dependent on the selected background.    
We then employ the counting methodology based on the $a$- and $c$-function, which are typically used to characterize the trace anomaly of conformal field theories in four dimensions.   We compute these in the enhanced one-loop approximation and demonstrate that the degrees of freedom decrease monotonically from six degrees in the UV regime.  This behavior is independent of the backgrounds considered.  
Further salient issues, such as the connection between the Fradkin--Tseytlin prescription and counting based on the $a$- and $c$-function, or the applicability of both methods in Einstein's gravity, are also addressed.
%

%
\end{abstract}

\maketitle

\section{Introduction}

A key aspect of the Renormalization Group (RG) is that quantum phenomena are 
described differently at different energy
scales. In particular, performing finite RG transformations and tracking the resulting flow allows one to connect descriptions of the system at distinct scales. 
The latter usually correspond to the inverse characteristic length scales, $\ell$, that are relevant in the physical description. 
The running energy scale $E=\ell^{-1}$, is a crucial parameter in the RG flow, akin to time in the description of hydrodynamic flow.  
Instead of $E$, a dimensionless logarithmic RG time is often used for convenience. This is typically defined as $t=\log({E}/E_{0})$, where $E_{0}$ is some reference energy scale.

It is now well recognized that due to this RG flow various quantities in quantum field theories (QFT's) necesarilly undergo changes and are different in UV (ultraviolet) and IR (infrared) regimes of a theory. Examples include interaction
couplings (the paradigmatic example being the fine-structure constant in QED~\cite{Anastasi}), wave-function renormalizations of quantum fields, anomalous dimensions of operators, or various Green functions, as they all change with the RG time $t$. Remarkably, the running exists even in QFT models that are completely scale-invariant at the classical level. In the latter case, in classical theory, the correlation functions do not depend on the energy $E$ and all coupling parameters are constant (that is, not running) and dimensionless. In addition to scale invariance, such models often have complete conformal symmetry at the classical level if the energy-momentum tensor has the appropriate form~\cite{Polchinski:1988}.
However, quantum effects typically break the classical
scale invariance (or enhanced conformal invariance) due to RG running. Only very special (constrained
and symmetric) models can remain scale invariant (or conformal) also at the quantum level~\cite{Fradkin:1985am}.

A particularly important quantity that may exhibit a dependence on the RG energy scale is the number of degrees of freedom (DOF). Here, DOF refer to the total number of polarization modes. Since Wilson's seminal work~\cite{Wilson}, it has become clear that RG transformations from UV to IR generally eliminate degrees of freedom associated with higher-energy modes. This is because in the passage from UV to IR, only the low-energy modes with $p^{2}\ll\Lambda_E^{2}$ (where $\Lambda_{ E}$ is the cutoff energy scale) are retained in the functional integrals in order to obtain the correct partition function or transition amplitudes at low energies.
Thus, in the vicinity of the IR fixed point, just the low energy
modes quantum fluctuate, propagate and are active, while the high
energy modes do not move and are fixed and passive. 
Consequently, for a given RG-transformation there might be no inverse, which is epitomized by a semigroup structure of RG. 

It is interesting to note that the change in the number of DOF along the RG flow from UV to IR has never been much studied in a systematic way in the literature. This striking absence is even bolstered by the fact that there is not unique agreed-upon definition of degrees of freedom in many QFT problems. This issue is particularly pressing, e.g., for condensed matter QFT systems such as non-Fermi liquids~\cite{Sachdev}, or quantum gravity systems such as higher derivative gravity theories~\cite{Fradkin:1985am}. It will be the higher derivative gravity that will be of interest here. In particular, 
%
we will examine two proposals for the number of DOF, namely a method based on the partition function, which to our knowledge first appeared in the paper of Fradkin and Tseytlin~\cite{Fradkin:1985am} (hence we call it the FT method), and a method based on the $c$- and $a$-anomaly~\cite{duff}.

The first proposal --- the FT method, is essentially based on a functional-integral approach.  The FT formula for DOF can be written as~\cite{Fradkin:1985am} 
\begin{equation}N_{{\rm DOF}} \ = \ \frac{\log Z^{2}}{\log Z_0^2}\, .
\label{ndofFT0}
\end{equation}
Here, $Z$ represents the partition function of the QFT of interest. The resulting expression is normalized with respect to the partition function $Z_0^2$ of the ancillary theory given by a  real massless Klein--Gordon scalar field, which serves as a reference degree of freedom.
The rationale behind this formula, as outlined in~\cite{Fradkin:1985am}, is as follows. For any linearized quantum gravity theory, the partition function can be put into the form $Z = Z_0^{\nu}$, where $Z_0 = \left(\det_0 \square\right)^{-1/2}$ and $\nu$ is the number of dynamical degrees of freedom described by the theory. 
Therefore, the desired result for the number of DOF is obtained by dividing the logarithms of the squares of these two partition functions. Heuristically one can also understand~(\ref{ndofFT0}) as a fraction of two free energies (when working in the Euclidean regime). Since free energy is an extensive state function, it must be proportional to the volume times the number of polarization DOF.  For  linearized gravity the proportionality constant is the same as for scalar massless field, hence the fraction should provide required number of DOF. 
The number of DOF obtained in this way basically corresponds to the virial degrees of freedom, since such DOF are related to the quadratic action. The quadratic action (namely its gradient part) dominates over the non-linear potential term at very high energies, i.e. in the UV regime. As the energy scale is lowered, the simple virial prescription will no longer hold and the number of DOF will flow. It should be emphasized that while~(\ref{ndofFT0}) is by definition formulated in a background independent manner, the actual computations in 4 dimensions are inevitably done in the framework of linearized gravity in some fixed background.   
%
%
In the following calculations we will use a perturbative approximation for the partition functions $Z$ to one-loop accuracy.
The issue that we will be particularly interested in is how the one-loop level DOF depend on the background used.

The second method, which exploits $a$- and $c$-anomaly, is based on the work of Zamolodchikov, who performed the earliest studies on the non-perturbative meaning of the number of DOF, specifically in 2-dimensional conformal field theories (CFT's)~\cite{key-1}. 
In this work, Zamolodchikov introduced the celebrated $c$-function, which interpolates
between two Virasoro central charges belonging to the limiting CFT's 
located at the corresponding UV and IR fixed points (FPs) of the RG.
In such CFT's there is no RG flow, no energy
dependence, and no changes with the scale, in particular, the number
of DOF is fixed, and it is (roughly) counted
by the central charge of the CFT. We remark here that the two CFT's
in the UV and IR regime must not be the same, hence also their central
charges are typically different, i.e. $ c_{{\rm UV}}\neq c_{{\rm IR}} $.
They may come with completely different spectra of primary conformal
operators, anomalous conformal dimensions, CFT data, etc. Moreover, the degrees of freedom between these two CFT's can be of completely different character, with different dynamics and different symmetry principles. 

The aforementioned connection with degrees of freedom suggests that the relation between the central charges of two limiting
CFT's should satisfy a sharp inequality 
\begin{equation}
c_{{\rm IR}} \ < \ c_{{\rm UV}}\, .
\label{eq: inequality}
\end{equation}
%
%
In fact, Zamolodchikov demonstrated that the inequality (\ref{eq: inequality}) holds for all 2-dimensional CFTs. Furthermore, he also provided an explicit construction for the interpolating $c$-function.
This celebrated $c$-function is a function of the energy scale $E$,
i.e. $c=c(E)$, which interpolates smoothly and monotonically along the RG flow between two central charges in the UV and IR regime~\cite{key-1}, so that
\begin{equation}
c(E\to0) \ = \ c_{{\rm IR}}\quad{\rm and}\quad c(E\to+\infty) \ = \ c_{{\rm UV}}\, .
\label{eq: limits_of_c}
\end{equation}
The $c$-function can be viewed as an interpolating running central charge that smoothly transits between UV and IR fixed points of the RG. It should,  however, be stressed that at non-extremal energies, a valid QFT description of the RG flow is typically not available since the CFT only describes the physics at fixed points of the RG. 
Thus, the interpolating $c$-function allows for an unperturbative definition of the {\em effective} DOF on energy scales between UV and IR FPs.  The monotonicity of $c$, expressed by the fact that $c'(E)>0$, is fully consistent with our intuition that the flow of DOF in any reasonable unitary local QFT should decrease monotonically with decreasing energy. This is because the RG flow is, in
a sense, always caused by the mathematical procedure of integrating
out massive large momentum modes. 
For these reasons, the monotonic decrease of the $c$-function is what was originally called the irreversibility of the RG flow between two CFT's in $d=2$ by Ref.~\cite{key-1}. 


Zamolodchikov's construction has 
been extended in various contexts, including higher dimensions \cite{key-3,key-4,key-5,key-6},
other theories \cite{key-7,key-8}, and different spacetime topologies
\cite{key-9,key-9b}, among others.
Such extensions led to the construction of $c$-functions, $a$-functions, $f$-functions, etc. and ensuing mathematical conjectures known as  $c$-, $a$- and $f$-theorem, respectively, cf.~e.g.~\cite{key-9,key-10,key-11,key-12,key-13}. These conjectures have both perturbative \cite{key-14} and also non-perturbative~\cite{key-15} formulations. Moreover, the knowledge obtained was instrumental in building the general theory of RG flows in
QFT~\cite{key-16, key-17, key-18, key-19}.
%
%
Of course, since a generic RG flow departs from quantum conformality, it is inevitably
related to the issue of conformal anomaly \cite{key-22, key-23, key-24} and
the way that conformal symmetry (or scale invariance) present at the FP, can be broken.

Our aim here is to study how the number of DOF in quantum (Weyl) conformal gravity (QCG) varies when moving away from the UV fixed point. There are several reasons for studying QCG. 
Although more than a century has passed since the introduction of Einstein's general theory of relativity (GR), GR, with its Ricci scalar-based action, has passed numerous experimental tests at solar-system scales and smaller. However, theoretical principles (namely the extension to the quantum domain) and recent large-scale cosmological observations challenge GR. This in turn has stimulated interest in various extensions of GR, as well as in alternative theories of gravity~\cite{Capozzilello,JL22,JL23}.
Among alternative gravitational theories, Weyl conformal gravity (WCG)~\cite{Bach} has recently received much attention as a strong candidate for explaining the dark matter problem.~\cite{Mannheim:89,Mannheim:12,Mannheim:12b}.  

WCG replaces the Einstein--Hilbert Lagrangian with the Weyl tensor squared Lagrangian, and in this respect, it belongs to the class of higher derivative (HD) gravity theories. 
It is now well recognized that  WCG is not just a simple generalization of the HD gravity models augmented with Weyl scale symmetry. 
In fact, the van~Dam--Veltman--Zakharov
discontinuity theorem~\cite{vanDam,Zakharov} ensures that no limiting procedure exists that would reduce generic HD theory to conformal gravity, both on the level of action and also on the level of physical observables. This brings 
a number of delicate technical and conceptual issues (particularly on the quantum level) that are largely dissimilar from those encountered in
generic HD quantum gravity models. 
Moreover, WCG is determined completely by a single dimensionless coupling constant, making WCG power-counting renormalizable and unique.
Despite a number of (seeming or real) pathologies, such as the appearance of perturbative ghost states~\cite{Riegert} or breakdown of scale invariance due to trace anomaly~\cite{RJ}, QCG provides interesting, conceptually rich, and largely
uncharted territory among HD gravity theories. 

Of particular interest here is the recent suggestion~\cite{JKS,JRGK} that QCG, after (global) conformal symmetry breaking, approaches the Starobinsky $f(R)$ model in its low energy phase.  In this connection, we may recall that the Starobinsky model was originally motivated by the one-loop correction to Einstein gravity~\cite{Starobinsky:80}. 
While WCG has 6 (on-shell) DOF~\cite{Riegert}  (massless
spin-2 graviton, massless spin-1 vector boson, and mass-less spin-2 ghost field), the Starobinsky gravity possesses only 3 (on-shell) DOF~\cite{whitt} (massless spin-2 graviton and scalaron (spin-0) field). In addition, in the low curvature limit, the scalaron field decouples, and the Starobinsky gravity transforms into Einstein--Hilbert gravity with two DOF from the spin-2 graviton.
It would certainly be interesting to confirm (or disprove) the flow of DOF in QCG from 6 to 3 or 2 but, unfortunately, the {\em one-loop} and {\em enhanced one-loop} calculations we employ in this paper are not enough to bridge the 
energy gap between the UV and IR regimes. Nevertheless, the calculations are quite instructive both from the point of view of the qualitative behavior of the running of DOF and of properties of the two methods used --- primarily their background (in-)dependence.

The paper is organized as follows: in the following section we briefly revisit some of the aspects of Weyl conformal gravity, which will be required in the main body of the text. In Sec.~\ref{Sec.II}, we employ the FT method in the context of QCG to perform an explicit one-loop computation of the running of DOF for the 4-dimensional maximally symmetric spaces (MSS) and Ricci-flat backgrounds. The results of these computations will turn out to be highly background dependent, implying that an alternative approach to DOF computation should be employed. 
In particular,  in Secs.~\ref{Sec.III} and~\ref{Ref.IV}, we analyze the running of DOF via $c$-anomaly and $a$-anomaly  by employing the MSS and Ricci-flat background. The running obtained is background-independent and logically consistent up to the accuracy of the enhanced one-loop approximation.
%
In Sec.~\ref{Sec.V} we outline the links between the two methods (i.e., FT and anomaly based methods), through study of how the effective action transforms under a rescaling of the metric by a constant factor. A brief summary of results and related discussions are provided in Sec.~\ref{Concl.}.
For the reader's convenience, we relegate some more technical issues to three appendices. 

\section{Quantum conformal gravity}\label{Sec.I}

For the sake of consistency, we will now briefly review the key aspects of QCG that will be needed in the following sections. More detailed exposition can be found, e.g. in Ref.~\cite{RJ}.

{The Weyl conformal gravity  is a pure metric theory that is invariant not only under the action of
the diffeomorphism group, but also under the Weyl rescaling of the metric tensor  by 
local smooth functions $\Omega(x)$: $g_{\mu\nu}(x)\rightarrow \Omega^2(x)g_{\mu\nu}(x)$.
The simplest WCG action functional in four spacetime dimensions that is both diffeomorphism and
Weyl-invariant has the form~\cite{Weyl1,Bach},
\begin{equation}
S[g] \ = \ -\frac{1}{4\alpha_{\rm{w}}^{\tiny{2}}}\int d^4x \ \! \sqrt{|g|} \ \! C_{\mu\nu\rho\sigma}
C^{\mu\nu\rho\sigma}\, ,
\label{PA1}
\end{equation}
where  $C_{\mu\nu\rho\sigma}$ is the
{Weyl tensor} which can be written as
\begin{eqnarray}
C_{\mu\nu\rho\sigma}  &=&
R_{\mu\nu\rho\sigma} \ - \ \left(g_{\mu[\rho}R_{\sigma]\nu} -
g_{\nu[\rho}R_{\sigma]\mu} \right)
\nonumber\\[1mm]
&+& \frac{1}{3}R \ \! g_{\mu[\rho}g_{\sigma]\nu}\, ,
\label{P2a}
\end{eqnarray}
with $R_{\mu\nu\rho\sigma}$ being the {Riemann curvature tensor},
$R_{\mu\rho}=R_{\mu\nu\rho}{}^\nu$  the  {Ricci tensor}, and $R= R_\mu{}^\mu$
the {scalar curvature}. Here and throughout, we employ the time-like metric signature $(+,-,-,-)$  whenever  pseudo-Riemannian (Lorentzian) manifolds are considered. 
The dimensionless coupling constant $\alpha_{\rm{w}}$ is chosen to mimic the Yang--Mills action. 
As for the notation for various scalar invariants (with four derivatives of the metric tensor), we accept the following conventions: for the square of
the Riemann tensor contracted naturally (preserved order of indices), that is $R_{\mu\nu\rho\sigma}R^{\mu\nu\rho\sigma}$, we use the symbol
$R^2_{\mu\nu\rho\sigma}$; the square of the Ricci tensor $R_{\mu\nu}R^{\mu\nu}$, we denote by simply $R_{\mu\nu}^2$; the square of the Ricci
curvature scalar is always $R^2$, while for the Weyl tensor square (with a natural contraction of indices) $C_{\mu\nu\rho\sigma}C^{\mu\nu\rho\sigma}$,
we employ a shorthand and schematic notation $C^2$. When the latter is treated as a local invariant (not under a volume integral, so without the possibility of
integrating by parts) in $d=4$ dimensions, one finds the following expansion of the $C^2$ invariant into standard invariants quadratic in curvature
\begin{equation}
C^2 \ = \ R_{\mu\nu\rho\sigma}^2 \ - \ 2R_{\mu\nu}^2 \ + \ \frac{1}{3}R^2\,.
\label{weylsquare}
\end{equation}
Finally, we will also need yet another important combination of the quadratic curvature invariants, namely Gauss--Bonnet (GB) term
\begin{equation}
E_4 \ = \ R_{\mu\nu\rho\sigma}^2 \ - \ 4R_{\mu\nu}^2 \ + \ R^2\, ,
\label{GBdef}
\end{equation}
which in $d=4$ is the integrand of the Euler–Poincaré invariant~\cite{Percacci}
\begin{eqnarray}
\chi \ = \ \frac{1}{32\pi^2}\int d^4x \ \! \sqrt{|g|} \ \! E_4\, .
\label{II.8.aa}
\end{eqnarray}
With the help of the Chern--Gauss--Bonnet  theorem, one
can cast the Weyl action $S$ into an equivalent form (modulo topological term)
\begin{equation}
S[g] \ = \
-\frac1{2\alpha_{\rm{w}}^2}\int d^4x\ \! \sqrt{|g|} \ \!\left(R_{\mu\nu}^2 \ - \ 
\frac{1}{3} R^2\right).
\label{PA2}
\end{equation}
It should be stressed that both (\ref{PA1}) and (\ref{PA2}) are Weyl-invariant only in $d=4$ dimensions.
In fact, under the Weyl transformation $g_{\mu \nu} \to \Omega^{2} {g}_{\mu \nu}$, the densitized square of the Weyl tensor
transforms as
\begin{eqnarray}
\sqrt{|g|} \ \! C^2 \ \to \ \Omega^{d-4} \sqrt{|g|}  \ \! C^2\, ,
\end{eqnarray}
in a general  $d$-dimensional spacetime. At the same time, the term $\sqrt{|{g}|} \ \!E_4$  supplies topological (and thus also Weyl) invariant only in $d=4$.} 
It should be emphasized that while the topological term is clearly unimportant at the classical level, it is relevant at the quantum level, where summation over different topologies should be considered. However, even if one stays with topologies with a fixed Euler--Poincaré invariant, the renormalization procedure will inevitably produce (already at one loop) the GB term~\cite{Fradkin:1985am}.  
Variation of $S$ with respect to the metric
yields the field equation (Bach vacuum equation)
\begin{equation}
\nabla_{\kappa} \nabla_{\lambda}C^{\mu\kappa\nu\lambda}_{\phantom{\lambda\mu\nu\kappa}}
\ -  \ \frac{1}{2} \ \! C^{\mu\kappa\nu\lambda}R_{\kappa\lambda} \ \equiv \ B^{\mu\nu}
\ = \  0\, ,
\label{Z42A}
\end{equation}
where $B^{\mu\nu}$  is the Bach tensor and $\nabla_{\alpha}$ is the usual covariant derivative. If, on a given background, $B^{\mu\nu}=0$, then the corresponding solution is known as the Bach-flat solution.

{To quantize WCG one formally defines the gravitational functional integral 
%
\begin{equation}
Z \ = \ \int_{\Sigma} {\cal D}g \ \! e^{iS[g]}\, .
\label{Z42Ab}
\end{equation}
Here, $g$ denotes ``geometries," that is, equivalent classes of metrics modulo diffeomorphisms and, in the case of QCG, Weyl transformations.   The space $\Sigma$ is the space of
all geometries. Although it is possible to give a mathematically precise definition of the space $\Sigma$, it is not easy to handle~\cite{Hamber}. Consequently, one typically employes the standard Fadeev--Popov procedure (which allows to define the functional integral directly on the space of metrics) and the background field method. 

To avoid problems related to the renormalization of non-physical sectors (i.e., Faddeev--Popov (FP) ghosts and longitudinal components of the metric field), it is
convenient  to use the York decomposition of the metric fluctuations
$h_{\mu\nu}$, defined as
\begin{eqnarray}
g_{\mu\nu}\ = \ g^{(0)}_{\mu\nu} \ + \ h_{\mu\nu}\,,
\end{eqnarray}
where we have denoted the background metric as $g^{(0)}_{\mu\nu}$. The York decomposition is then implemented in two steps~\cite{JRGK,Percacci}. In the first step the metric fluctuations are rewritten as
%
\begin{eqnarray}
h_{\mu\nu}\ = \ \bar{h}_{\mu\nu} \ + \ \frac{1}{4}g_{\mu\nu}h\, ,
\end{eqnarray}
where $h$ is a trace part of $h_{\mu\nu}$ and  $\bar{h}_{\mu\nu}$ is the ensuing traceless part.
More specifically,
\begin{eqnarray}
&&g^{(0)\mu\nu}\bar{h}_{\mu\nu} \ = \  \bar{h}_{\mu}{}^{\mu}\   = \  0\, , \nonumber \\[1mm]\quad
&&h \ = \  g^{(0)\mu\nu}h_{\mu\nu} \  = \  h_{\mu}{}^{\mu}\, .
\end{eqnarray}
In our subsequent derivations, it will always be implicit that the Lorentz indices are raised or lowered
via the background metric, i.e. via $g^{(0)\mu\nu}$ or $g_{(0)\mu\nu}$ respectively.
Also all covariant derivatives
$\nabla_\mu$ below will be understood as taken with respect the background metric. By the symbol $\square$ we denote the covariant box operator defined as $\square\equiv\nabla^\mu\nabla_\mu$.

In the second step, one decomposes the traceless part into the transverse, traceless
tensor $\bar{h}_{\mu\nu}^\perp$  and to parts carrying the longitudinal degrees of freedom, namely
\begin{eqnarray}
\bar{h}_{\mu\nu} &=& \bar{h}_{\mu\nu}^{\perp}\  + \ \nabla_{\mu}\eta_{\nu}^{\perp}\  + \ \nabla_{\nu}\eta_{\mu}^{\perp}\nonumber \\[1mm]
&&+ \  \nabla_{\mu}\nabla_{\nu}\sigma \  - \ \frac{1}{4}g_{\mu\nu} \ \!\Box \ \!\sigma\, .
\label{Ydecomp}
\end{eqnarray}
These mixed-longitudinal (and traceless) parts are written in terms of an arbitrary transverse vector
field $\eta_\mu^\perp$ and a scalar (trace) degree of freedom $\sigma$. The latter fields must
satisfy the transversality and  tracelessness conditions, i.e.
\begin{eqnarray}
\nabla^{\mu}\bar{h}_{\mu\nu}^{\perp}\ = \ 0,\quad\nabla^{\mu}\eta_{\mu}^{\perp} \ = \ 0,
\quad\bar{h}_{\,\,\mu}^{\perp\mu} \ = \ 0\, .
\end{eqnarray}
The true physical field in QCG is the transverse and traceless field $\bar{h}_{\mu\nu}^{\perp}\equiv h^{TT}_{\mu\nu}$. Indeed, from the second variation of the
Weyl action  expanded around a generic background it can be seen that $\bar{h}_{\mu\nu}^{\perp}$ is the only field component that propagates on quantum level.
The vector field $\eta_\mu^\perp$ and two scalar fields $h$ and $\sigma$ completely drop out from the expansion due to diffeomorphism and conformal invariance, respectively.

In order to compute the partition function, it is important to first set up a notation regarding the functional determinants and ensuing functional traces.
The character of the space in which these determinants (and related traces) are to be evaluated is denoted by the subscript immediately after the symbol ``det'', while the superscript always denotes the standard power.
The corresponding internal traces in the matrix space of fluctuations actually count the field-theoretic
number of DOF in such subspaces if they are taken from the unity $\hat{\mathbb{1}}$.
We will denote such traces by the symbol ``tr''. We stress that the functional traces, denoted by ``Tr'', are different because they contain also the
integrations over the background spacetime.  There are few subspaces of fluctuations in which we would like to consider our determinants.
They mainly depend on the spin of the fluctuations and whether they are transverse,
traceless or completely unconstrained fields. We have a description of various fields in subscripts:
\begin{itemize}
\item $0$ -- spin-0 scalar field with 1 DOF, that is ${\rm tr}_{0}\hat{\mathbb{1}}=1$.
\item $1\!\perp\,\equiv1T$ -- spin-1 constrained vector field $v_{\mu}^{T}\equiv v_\mu^\perp$ to be transverse $\nabla^{\mu}v_{\mu}^{T}=0$ with 3 DOF, that is ${\rm tr}_{1\perp}\hat{\mathbb{1}}=3$.
\item $1$ -- spin-1 unconstrained vector field $v_{\mu}$ with 4 DOF, that is ${\rm tr}_{1}\hat{\mathbb{1}}=4$.
\item $2\!\!\perp\,\equiv2TT$ -- spin-2 fully constrained tensor rank-2 symmetric field $h_{\mu\nu}^{TT}\equiv\bar h_{\mu\nu}^\perp$, with a condition to be
transverse and traceless $h^{TT}{}_\mu{}^\mu=\nabla^{\mu}h_{\mu\nu}^{TT}=0$ with 5 DOF, that is ${\rm tr}_{2\perp}\hat{\mathbb{1}}=5$.
\item $2T$ -- spin-2 partially constrained tensor rank-2 symmetric field $h_{\mu\nu}^{T}\equiv \bar h_{\mu\nu}$, with a condition to be only traceless
$h^{T}{}_\mu{}^\mu=0$ with 9 DOF, that is ${\rm tr}_{2T}\hat{\mathbb{1}}=9$.
\item $2$ -- spin-2 fully unconstrained tensor rank-2 symmetric field $h_{\mu\nu}$ with 10 DOF, that is ${\rm tr}_{2}\hat{\mathbb{1}}=10$.
\end{itemize}
Note in particular the differences in the use of the superscript ``T'' for spin-1 and spin-2 fields.



\section{Fradkin--Tseytlin formula for DOF's in Conformal Weyl Gravity \label{Sec.II}}

In this section, we are going to use the FT prescription for the effective number of field-theoretic DOF's to calculate how the DOF's will run in QCG. Our computation will be done for two classes of Bach-flat backgrounds --- 4-dimensional MSS and a Ricci-flat background.

%
\subsection{Flow on a MSS background \label{Sec.IIIA}}

On a 4-dimensional MSS manifold (like a 4-sphere) we have the following curvature relations
\begin{eqnarray}
 &&R_{\mu\nu} \ = \ \Lambda g_{\mu\nu}\,,\nonumber \\[2mm]
 &&R_{\mu\nu\rho\sigma} \ = \  \frac{\Lambda}{3} \left(g_{\mu\rho} g_{\nu\sigma}- g_{\mu\sigma} g_{\nu\rho} \right),\nonumber \\[2mm]
 &&R \ = \ 4\Lambda\,,\nonumber \\[2mm]
 && R^2= 16\Lambda^2\,,\nonumber \\[2mm]
 &&R_{\mu\nu}^2= 4\Lambda^2\, ,
\end{eqnarray}
and the Kretschmann invariant scalar
\begin{eqnarray}
 R_{\mu\nu\rho\sigma}^2 \ = \  K \ = \ \frac83\Lambda^2\,,
\end{eqnarray}
so, that one can substitute
\begin{eqnarray}
 \Lambda \ = \ \sqrt{\frac{3K}{8}}\, .
\end{eqnarray}

Here $\Lambda$ is a curvature parameter characterizing completely the MSS background and it is related to the cosmological constant parameter by the gravitational equation of motion. Actually, in WCG, all the MSS backgrounds for any value of $\Lambda$ are vacuum solutions, without the need of introducing of the cosmological constant.
%

The square of the one-loop partition function of QCG theory on a 4D sphere characterized
by $R=4\Lambda$ (Ricci scalar) can be put into form~\cite{RJ}
\begin{eqnarray}
&&\mbox{\hspace{-4mm}}Z_{{\rm 1-loop}}^{2}\nonumber \\[2mm]
&&\mbox{\hspace{-4mm}}= \ \frac{\det_{1}^{2}(\square+\Lambda)\det_{1}\left(\square+\frac{1}{3}\Lambda\right)\det_{0}\left(\square+\frac{4}{3}\Lambda\right)}{\det_{2T}\left(\square-\frac{2}{3}\Lambda\right)\det_{2T}\left(\square-\frac{4}{3}\Lambda\right)\det_{0}(\square+2\Lambda)}\,. ~~~~~~~
\label{21.cf}
\end{eqnarray}
Strictly speaking, all determinants in the above formula should be computed from the dimensionless operators, in particular, all operators $(\square+\chi\Lambda)$ (where $\chi$ is a numerical constant) must be divided by the square of some energy scale.  With this proviso, Eq.~(\ref{21.cf}) should be rewritten as
\begin{eqnarray}
&&\mbox{\hspace{-4mm}}Z_{{\rm 1-loop}}^{2}\nonumber \\[2mm]
&&\mbox{\hspace{-4mm}}= \ \frac{\det_{1}^{2}\left(\frac{\square+\Lambda}{\mu^2}\right)\det_{1}\left(\frac{\square+\frac{1}{3}\Lambda}{\mu^2}\right)\det_{0}\left(\frac{\square+\frac{4}{3}\Lambda}{\mu^2}\right)}{\det_{2T}\left(\frac{\square-\frac{2}{3}\Lambda}{\mu^2}\right)\det_{2T}\left(\frac{\square-\frac{4}{3}\Lambda}{\mu^2}\right)\det_{0}\left(\frac{\square+2\Lambda}{\mu^2}\right)}\,. \label{formula22}~~~~~\end{eqnarray}
Similarly, the FT formula for the number of DOF takes the form
\begin{equation}
N_{{\rm DOF}} \ = \ -\frac{\log Z_{{\rm 1-loop}}^{2}}{{\log\det_{0}\left(\square/\mu^2\right)}}\, ,
\label{ndofFT}
\end{equation}
where both in (\ref{formula22}) and (\ref{ndofFT}) we have introduced the renormalization sliding scale $\mu$, so that $\lambda={\Lambda}/{\mu^{2}}$ is a dimensionless parameter. With this, we can write
\begin{eqnarray}
&&\mbox{\hspace{-7mm}}N_{{\rm DOF}} \ = \ N_{{\rm dof}}(\lambda)=-\frac{\log Z_{{\rm 1-loop}}^{2}}{\log\det_{0}\left(\frac{\square}{\mu^{2}}\right)}\nonumber \\[2mm]
&=&-\frac{2\log\det_{1}\left(\frac{\square}{\mu^{2}}+\lambda\right)+\log\det_{1}\left(\frac{\square}{\mu^{2}}+\frac{1}{3}\lambda\right)}{\log\det_{0}\left(\frac{\square}{\mu^{2}}\right)}\nonumber \\[2mm]
&&+ \ \frac{-\log\det_{0}\left(\frac{\square}{\mu^{2}}+\frac{4}{3}\lambda\right) + \log\det_{2T}\left(\frac{\square}{\mu^{2}}-\frac{2}{3}\lambda\right)}{\log\det_{0}\left(\frac{\square}{\mu^{2}}\right)} \nonumber \\[2mm]
&&+ \ \frac{\log\det_{2T}\left(\frac{\square}{\mu^{2}}-\frac{4}{3}\lambda\right)+\log\det_{0}\left(\frac{\square}{\mu^{2}}+2\lambda\right)}{\log\det_{0}\left(\frac{\square}{\mu^{2}}\right)}\, ,~~~~~~
\label{ndof0}
\end{eqnarray}
which is a  finite expression. In the UV regime, when $\mu^{2}\gg\Lambda$, we have $\lambda\ll1$,
so based on \eqref{ndof0}, we reproduce flat spacetime tree-level result with
\begin{eqnarray}
N_{{\rm DOF}}(\lambda=0)&=&-(2\times4+4+1)+9+9+1
\nonumber\\[2mm]
&=& -13+19 \ = \ 6\, .\label{formula25}
\end{eqnarray}
%
%
Now the question stands, how does (\ref{ndof0}) runs when changing the $\lambda$ parameter or the $\mu$ scale?
Take, for example, a determinant of the first scalar term  $
\frac{\square}{\mu^{2}}+\frac{4}{3}\lambda$ in the numerator of \eqref{formula22}.
In order to tackle ``$\log{\det}$'' term, we shall use the heat kernel method with the FT regularization prescription, namely
\begin{eqnarray}
&&\mbox{\hspace{-10mm}}\log{\det}_0\left(\frac{\square}{\mu^{2}}+\frac{4}{3}\lambda\right) \ = \ {\rm Tr}\log\left(\frac{\square_0}{\mu^{2}}+\frac{4}{3}\lambda\mathbf{1}_0\right) \nonumber \\[2mm]
&&= -\int_{\epsilon}^{\infty}\frac{ds}{s}\ \!{\rm Tr}\exp\left[-s\left(\frac{\square_0}{\mu^{2}}+\frac{4}{3}\lambda\right)\right]\, ,\label{hk26}
\end{eqnarray}
where $\epsilon \ = \ \left({\mu}/{L}\right)^{2}$ and where $L$ is an UV-cutoff energy scale.
%
%
In this case, ``Tr'' denotes the functional trace, i.e. the integration over the
the background spacetime.
Analogous computation  can be done for other ``$\log{\det}$'' terms. For determinants of non-scalar terms, ``Tr'' would also include the trace over internal degrees of freedom. 
Without the regulator $\epsilon$, the integral in~(\ref{hk26}) diverges at the lower limit (i.e. for small dimensionless proper time $s$). Since small proper time corresponds to small distances $(x-y)\sim \tau^{1/2}$ or equivalently to large momenta $k \sim \tau^{-1/2}$, the corresponding divergence is an UV type. The dimensionless version of $\tau$, i.e. $\epsilon$, can thus be chosen as $\epsilon = (\mu/L)^2$, where $\mu$ can be identified with the sliding energy scale. In particular, $\mu$ roughly measures the distance from the UV fixed point, i.e. for $\epsilon=0$ we are at the UV fixed point and for $\epsilon \sim 1$ at the IR fixed point.



%
In the FT regularization, the lower bound of the integration in (\ref{hk26}) is $s_-=\epsilon$, so when the UV-cutoff $L$ is completely removed (i.e., $L\to+\infty$) the integration produces UV-divergences for the partition function $Z$. However, as we will show shortly, these divergences between numerator and denominator cancel out in~(\ref{ndofFT}), leaving a final result for the number of DOF.

In the general case, the corresponding heat kernel coefficients are defined via the Seeley--De~Witt expansion
\begin{eqnarray}
&&\mbox{\hspace{-12mm}}{\rm Tr}\exp\left[-s\left(\frac{\square_0}{\mu^{2}}+\chi\lambda\right)\right] \nonumber \\[2mm]  &&= \ \frac{1}{(4\pi)^2}\int d^4x \ \! \sqrt{|g|}\ \!\sum_{p=0}^{\infty}\left(\frac{s}{\mu^{2}}\right)^{(p-4)/2}b_{p}'\, ,\label{hk27}
\end{eqnarray}
where the above formula is understood as the asymptotic series expansion near $s\to 0^{+}$ ($s\ll1$)
and similarly we have for the determinant in the denominator  of~(\ref{ndof0})
\begin{eqnarray}
&&\mbox{\hspace{-10mm}}{\rm Tr}\exp\left[-s\left(\frac{\square_0}{\mu^{2}}\right)\right]  \nonumber \\[2mm] 
&&= \ \frac{1}{(4\pi)^2}\int d^4x \ \!\sqrt{|g|} \ \! \sum_{p=0}^{\infty}\left(\frac{s}{\mu^{2}}\right)^{(p-4)/2}b_{p}\, .\label{hk28}
\end{eqnarray}
In our case only even positive integer values of the summation index $p$ matter. Moreover, for $p\geqslant4$ we extract from the integrals in (\ref{hk26}) only parts with the most leading $\epsilon$-dependence in the UV regime. This is to avoid problems in the IR limit of the $s$-integration (large values of the $s$ parameter).

In the case of the scalar representation for the box operator and for $\chi=\frac{4}{3}$ the  heat kernel coefficients $b_p$ and $b_p'$  read~\cite{Kluth:2019vkg,RJ}
\begin{eqnarray}
b_{0} &=& b'_{0} \ = \ 1\, ,\nonumber \\[2mm]
b_{2} &\equiv& b_{2}\left(\frac{\square_0}{\mu^{2}}\right) \ = \frac{1}{6} R = \ \mu^{2}\frac{1}{6}\frac{R}{\mu^{2}}\nonumber \\[2mm]
&=& \mu^{2}\frac{2}{3}\frac{\Lambda}{\mu^{2}} \ = \ \frac{2}{3}\mu^{2}\lambda\, , \nonumber \\[2mm]
b'_{2} &\equiv& b_{2}\left(\frac{\square_0}{\mu^{2}}+\frac{4}{3}\lambda\right) \nonumber \\[2mm] &=& \ \frac{2}{3}\mu^{2}\lambda+\frac{4}{3}\mu^{2}\lambda \ = \ 2\mu^{2}\lambda\, , \nonumber \\[2mm]
b_{4}&\equiv& b_{4}\left(\frac{\square_0}{\mu^{2}}\right) = \frac{29}{135}\Lambda^{2} \ = \ \frac{29}{135} \mu^4 \lambda^2\, ,\nonumber\\[2mm]
b'_{4}&\equiv& b'_{4}\left(\frac{\square_0}{\mu^{2}}+\frac43\lambda\right) \ = \ \frac{29}{135}\Lambda^{2}+\frac{8}{9}\Lambda^{2}+\frac{8}{9}\Lambda^{2}\nonumber \\[2mm]
&=&
\frac{269}{135}\Lambda^{2} \ = \ \frac{269}{135} \mu^4 \lambda^2\, .
\end{eqnarray}
For the un-primed $b_p$ coefficients, we have more generally 
\begin{eqnarray}
b_i \ \equiv \  b_i\left(\frac{\square_0}{\mu^2}\right),
\end{eqnarray} 
and since they always appear in the denominators of expressions like \eqref{ndofFT}, then these are defined for the specific scalar (spin-0) representation. Hence, for all the computations below, we have $b_0=1$, $b_2=\frac23 \mu^2\lambda$ and $b_4=\frac{29}{135}\mu^4\lambda^2$.
Now, the situation is different for the coefficients $b'_p$, because they depend explicitly on the shift number $\chi$ and also on the representation in which the box operator is considered. With full generality, then we defined
\begin{eqnarray}
b'_i \ \equiv \ b_i\left(\frac{\square}{\mu^2}+\chi\lambda\right)\,.
\end{eqnarray}
For the UV-divergent part (arising in the limit $\epsilon\to0$) of the functional traces of the general operator $\Delta=\frac{\square}{\mu^{2}}+\chi\lambda$, we find that

\begin{eqnarray}
&&\left({\rm Tr}\log\frac{\Delta}{\mu^{2}}\right)_{\infty} \nonumber \\[2mm] 
&&\mbox{\hspace{5mm}}= \ -\frac{1}{(4\pi)^2}\left(\frac{1}{2}L^{4}B'_{0}+L^{2}B'_{2}+\log\frac{L^{2}}{\mu^{2}}B'_{4}\right),~~~~~~~
\label{uvpart}
\end{eqnarray}
where $B'_p$ coefficients denote integrated coefficients of heat kernel expansion, i.e. $B'_p=\int d^4x\sqrt{|g|}\ \!b'_p$.

Returning to the case of the scalar representation for the box
operator and for $\chi=\frac43$, the corresponding contribution to $N_{\rm dof}$ thus reads (in the formula \eqref{ndof0} this comes with the overall minus sign)
%
\begin{eqnarray}
&&\mbox{\hspace{-6mm}}\frac{\log\det_{0}\left(\frac{\square}{\mu^{2}}+\frac{4}{3}\lambda\right)}{\log\det_{0}\left(\frac{\square}{\mu^{2}}\right)}\nonumber\\[2mm]
&&= \ \frac{\frac{1}{2}L^{4}b'_{0}+L^{2}b'_{2}+\log\frac{L^{2}}{\mu^{2}}b'_{4}-\frac{1}{L^{2}}b'_{6}+\ldots}{\frac{1}{2}L^{4}b{}_{0}+L^{2}b{}_{2}+\log\frac{L^{2}}{\mu^{2}}b{}_{4}-\frac{1}{L^{2}}b{}_{6}+\ldots}\,.~~~~~~
\label{11.vb}
\end{eqnarray}
%
%
%
This might be more explicitly written as 
\begin{eqnarray}
&&\mbox{\hspace{-4mm}}\frac{\frac{1}{2}L^{4}b'_{0}+L^{2}b'_{2}+\log\frac{L^{2}}{\mu^{2}}b'_{4}-\frac{1}{L^{2}}b'_{6}+\ldots}{\frac{1}{2}L^{4}b{}_{0}+L^{2}b{}_{2}+\log\frac{L^{2}}{\mu^{2}}b{}_{4}-\frac{1}{L^{2}}b{}_{6}+\ldots}\nonumber \\[2mm]
&&\mbox{\hspace{-4mm}}= \
\frac{\frac{1}{2}L^{4}+\gamma'_{2}\lambda\mu^{2}L^{2}+\gamma'_{4}\lambda^{2}\mu^{4}\log\frac{L^{2}}{\mu^{2}}-\gamma'_{6}\lambda^{3}\mu^{6}\frac{1}{L^{2}}+\ldots}{\frac{1}{2}L^{4}+\gamma{}_{2}\lambda\mu^{2}L^{2}+\gamma_{4}\lambda^{2}\mu^{4}\log\frac{L^{2}}{\mu^{2}}-\gamma_{6}\lambda^{3}\mu^{6}\frac{1}{L^{2}}+\ldots}\nonumber \\[2mm]
&&\mbox{\hspace{-4mm}}= \ \frac{\frac{1}{2}+\gamma'_{2}\lambda\frac{\mu^{2}}{L^{2}}+\gamma'_{4}\lambda^{2}\frac{\mu^{4}}{L^{4}}\log\frac{L^{2}}{\mu^{2}}-\gamma'_{6}\lambda^{3}\frac{\mu^{6}}{L^{6}}+\ldots}{\frac{1}{2}+\gamma{}_{2}\lambda\frac{\mu^{2}}{L^{2}}+\gamma_{4}\lambda^{2}\frac{\mu^{4}}{L^{4}}\log\frac{L^{2}}{\mu^{2}}-\gamma_{6}\lambda^{3}\frac{\mu^{6}}{L^{6}}+\ldots}\, .~~~~~~~~~
\label{Ndof1}
\end{eqnarray}
We note that for MSS we can factor out the integration over the 4-dimensional spacetime volume element, i.e. $\int d^4x\sqrt{|g|}$, and formally cancel it out as a ``common factor'' between the numerator and the denominator in (\ref{ndofFT}). This is possible because on any MSS any curvature invariant shows no spacetime point dependence, as these spacetimes are homogeneous. 

In~(\ref{Ndof1}), all $\gamma$'s and $\gamma'$'s are pure adimensional numbers (independent of $\mu$, $L$, $\lambda$, $\Lambda$). Since, in general, the coefficients $b_p$ (and similarly $b'_p$) have the following $\mu$-dependence: 
$b_p=\gamma_p \mu^p \lambda^{p/2}$, one can easily understand the general structure of the series both in the numerator and denominator of the last line of~\eqref{Ndof1}. Except for the first three terms in the denominator (as well as the numerator) of~\eqref{Ndof1}, which are special (as exemplified by the logarithmic term), the general term coming from $b_p$ for $p\geqslant6$ will be $-\frac{2}{p-4}\gamma_p\lambda^{p/2}\left(\frac{\mu}{L}\right)^p$. 

In terms of the $\epsilon$ variable, (\ref{Ndof1}) has the structure
\begin{eqnarray}
\frac{f'(\epsilon,\lambda)}{f(\epsilon,\lambda)}\, ,
\end{eqnarray}
where $f'$ is the same function as $f$ with $\gamma'$'s in place of  $\gamma$'s. Explicitly, in the example considered here, it is
\begin{eqnarray}
&&\mbox{\hspace{-6mm}}\frac{\log\det_{0}\left(\frac{\square}{\mu^{2}}+\frac{4}{3}\lambda\right)}{\log\det_{0}\left(\frac{\square}{\mu^{2}}\right)}\nonumber\\[2mm]
&&=\ \frac{\frac{1}{2}+\gamma'_{2}\lambda\epsilon-\gamma'_{4}\lambda^{2}\epsilon^2\log\epsilon-\gamma'_{6}\lambda^{3}\epsilon^3+\ldots}{\frac{1}{2}+\gamma_{2}\lambda\epsilon-\gamma_{4}\lambda^{2}\epsilon^2\log\epsilon-\gamma_{6}\lambda^{3}\epsilon^3+\ldots}\,.~~~~~~
\label{11.vb2}
\end{eqnarray}
%
From~(\ref{11.vb2}) it is clear that the effective number of degrees of freedom $N_{\rm DOF}$ (which is the dimensionless number) depends only on two dimensionless parameters, namely $\epsilon$ and the MSS curvature parameter $\lambda$.

Explicitly, we have for all the coefficients $\gamma$ and $\gamma'$, in the example of the term considered here
\begin{eqnarray}
    \gamma_2'\ = \ 2\, ,\quad \gamma_2 \ = \ \frac{2}{3}\, ,\quad \gamma_4'\ = \ \frac{269}{135}\, ,\quad\gamma_4 \ = \ \frac{29}{135}\, ,~~~~~
\end{eqnarray}
where we have restricted our attention only up to terms with $p\leqslant 4$ in (\ref{hk27}) and (\ref{hk28}).
A similar behavior holds also for the remaining 5 terms in (\ref{ndof0}).
 %

%

%
By inspection of the formula (\ref{Ndof1}), one can observe that the dominant contribution to the running comes from $b_2 $ and $b_2'$ coefficients, but for the sake of computation accuracy we will also keep the logarithmic terms, i.e.  terms proportional to $b_4$ and $b_4'$ . Although not pursued here, it is possible to include even higher order $b_p$'s. In such a case one can, for example, use the general heat kernel trace techniques on spheres as developed in~\cite{Kluth:2019vkg}.

By collecting the results for all six terms in~(\ref{ndof0}), we obtain, after some algebra, that the resulting effective number of DOF is given by
 \begin{eqnarray}
N_{\rm DOF}&=&\frac{6-\frac{136}{3}\frac{\Lambda}{L^{2}}+\frac{232}{5}\frac{\Lambda^{2}}{L^{4}}\log\frac{\mu}{L}}{1+\frac{4}{3}\frac{\Lambda}{L^{2}}-\frac{116}{135}\frac{\Lambda^{2}}{L^{4}}\log\frac{\mu}{L}}\label{form11}\, .
\end{eqnarray}
For reader's reference, all the necessary results for the coefficients $\gamma'_2$ and $\gamma'_4$  for all six terms of \eqref{ndof0} are presented in Ref.~\cite{RJ} and also can be found in appendix \ref{Appendix A}.

Note that 
in the limit $\lambda\to0$ (flat spacetime background), we obtain 
\begin{eqnarray}
\frac{f'(\epsilon,\lambda)}{f(\epsilon,\lambda)}\ \to \ \frac{\frac{1}{2}}{\frac{1}{2}}\ \to \ 1\, ,
\end{eqnarray}
for the case of the example scalar operator $\frac{\square}{\mu^{2}}+\frac{4}{3}\lambda$. This is true for all other five terms in \eqref{ndof0} as well, with the general results that in this limit the $b'_0$ coefficients for corresponding operators are attained. These $b'_0$ coefficients just count the number of DOF's of simple box operator in the particular representation equivalent to the dimension of its internal space. Thus, in this limit, the argument is exactly the same as in the formula \eqref{formula25}, which, as expected, results in 6 DOF, as on a flat spacetime background.

Let us now rewrite~(\ref{form11}) in terms of the $\epsilon$ variable as
 \begin{eqnarray}
N_{\rm DOF}&=&\frac{6-\frac{136}{3} \lambda \epsilon +\frac{232}{10}\lambda^2\epsilon^2\log\epsilon}{1+\frac{4}{3}\lambda\epsilon-\frac{116}{270}\lambda^2\epsilon^2\log\epsilon}\, \label{form0041}.\end{eqnarray}
The plot of this running for various values of the dimensionless parameter $\lambda$ is depicted in Fig.~\ref{Fig.1} as a function of the dimensionless ratio $\mu/L=\sqrt\epsilon$. For each curve there, we assume that the dimensionless parameter $\lambda$ is kept fixed and we vary only the values of the dimensionless $\epsilon$ parameter in the range $0\leqslant\epsilon\leqslant\frac{1}{25}= 0.04$. Although, the theoretically possible and maximal value of the $\epsilon$ is 1, when $\mu=L$ (i.e. when the sliding scale equals to the IR scale), we know that using perturbative calculus (here to the one-loop order)  we can trust our computation only in the vicinity of UV fixed point, so only for $\epsilon$ not so big. For explicitness, we choose $\epsilon<0.04$.
\begin{figure}[!ht]
\begin{center}
$\mbox{\hspace{4mm}}$\includegraphics[width= 8.0cm]{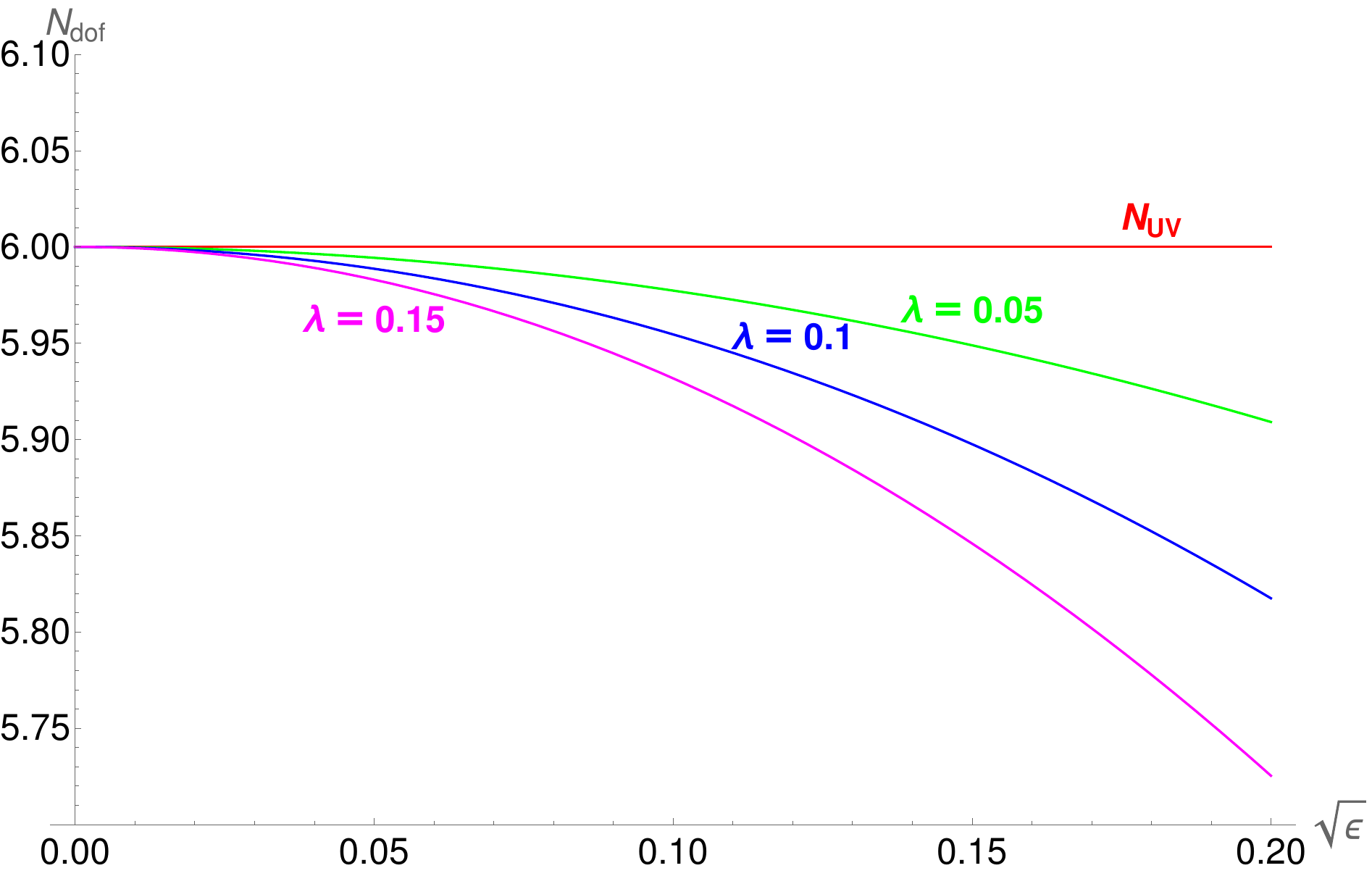}
\vspace{1mm}
\caption{Plot of $N_{\rm DOF}$ as a function of $\sqrt{\epsilon}=\mu / L$ for various positive values of $\lambda$. In red we show the value of DOF in the UV limit. Only the numerator of (\ref{form0041}) is considered. Thus, if $\lambda$ (or equivalently $\Lambda$) is sufficiently small,  DOF will decrease toward the IR regime.}
\label{Fig.1}
\end{center}
\end{figure}

As can be seen from Fig.~\ref{Fig.1}, the number of degrees of freedom in the UV regime reaches the value 6 for any value of the curvature parameter on MSS, i.e. for any $\lambda$. Clearly, the UV regime corresponds to the $L\to+\infty$ limit, i.e. to the $\epsilon\to0^{+}$ limit. Thus, when $\epsilon$ parameter is lowered to minimal value zero, the effective number of DOF's grows. On the other hand,  the number of DOF's decreases as soon as the $\epsilon$ parameter departures from the UV regime.

{
If we take in~(\ref{form0041}) the derivative with respect to $\epsilon$ at $\epsilon = 1/25$, we get
\begin{eqnarray}\label{form20}
&&\mbox{\hspace{-19mm}}\left. \frac{d N_{\rm DOF}}{d\epsilon}\right|_{\epsilon = 1/25}\nonumber \\[2mm] 
&&\mbox{\hspace{-14mm}}= \ \frac{1125000  \lambda \ \![-337500 + 29 \ \!\lambda \ \!(225+ 4 \lambda) ]}{[84375 + 4 \ \! \lambda \ \! (1125 + 29\lambda \log5)]^2}\nonumber \\[2mm] 
&&\mbox{\hspace{-9mm}}- \ 
\frac{1125000  \lambda \ \![ 29 \ \!\lambda  \ \!  (900 + 8\lambda) \log 5]]}{[84375 + 4 \ \!\lambda \ \! (1125 + 29\lambda \log5)]^2}
\, , \label{form13}
\end{eqnarray}
which is positive for $\lambda \in (-10.278,0)$ and $\lambda \gtrsim -127.57$. In such cases, DOF will increase (rather than decrease)
toward the IR regime. This is certainly an undesirable feature of the method, which, especially for large $|\lambda|$, may be due to the fact that our one-loop perturbation treatment is unwarranted in such a regime.
}

%
%

In passing, we might notice that it is also possible to neglect the curvature dependence of the denominators in the formula (\ref{ndofFT}), which are just the determinants of a single scalar field that is put on the same curved background (as the fields in the numerators live in) and minimally coupled to it. We can, however, choose to normalize~(\ref{ndofFT}) by the logarithms of the partition function of the single scalar field on the flat spacetime.
In such a case the constant $\frac{1}{2}L^4 b_0(\square_0/\mu^2) = \frac{1}{2}L^4$. In this way, we just analyze the numerator of the expression in \eqref{form11}, exactly with concordance of what was plotted in Fig.~\ref{Fig.1}. This would provide yet another (and simpler) definition of the effective number of DOF. 

 \subsection{Flow on Ricci-flat backgrounds \label{Sec.IIIB}}

In this subsection, we will focus on running of DOF in a Ricci-flat background.
We recall that Ricci-flat manifolds 
are defined so that their Ricci tensor 
 $R_{\mu\nu}=0$ and thus $R=0$. This in turn implies that the Riemann curvature tensor is the same as the completely traceless Weyl tensor, i.e. $R_{\mu\nu\rho\sigma}=C_{\mu\nu\rho\sigma}$, and then we have that $R_{\mu\nu\rho\sigma}^{2}=C^{2}= K=E_4$, where $E_4$ is an Euler invariant in four dimensions (also known as the Gauss--Bonnet density term). 
 %
 The scalar field $K$ is the so-called Kretschmann scalar  and on the whole manifold partially characterizes the Ricci-flat background. Actually, in WCG, all the Ricci-flat backgrounds
for any field profile $K$ are Bach flat.
As proved in~\cite{RJ}, for any Ricci-flat spacetime to be non-trivial (non-flat), it is required that the Kretschmann scalar $K$ cannot be a constant scalar field through the whole manifold. If Einstein space has a uniform Kretschmann scalar, then it must necessarily be MSS. Conversely, for Einstein space to be Ricci-flat, the cosmological parameter must vanish and the Kretschmann scalar must be non-trivial. As we will see, these facts will be important when trying to compute space-time-volume integrated coefficients in the heat kernel expansion.

To obtain the number of the gravitational DOF in WCG, we again use 
the FT defining relation
\begin{eqnarray}
N_{\rm DOF}=-\frac{\log Z^{2}}{\log\det(\square_{0}/\mu^2)}\, ,
\label{formula422}
\end{eqnarray}
where now on a generic Ricci-flat background~\cite{RJ}
\begin{eqnarray}
Z^{2}=\frac{\det^{3}\left(\square_{1}/\mu^2\right)}{\det^{2}\left(\frac{{\square}_{2T}+2\hat{C}}{\mu^2}\right)}\, .
\label{formula42}
\end{eqnarray}
On a Ricci-flat background, we will only perform calculations to the accuracy of $b_{4}$, since the results for higher $b_{2n}$ are generally not known. Moreover, up to this level,
only $b_{4}$ depends on the $\hat{C}$ tensor (matrix-valued form of the Weyl tensor $C_{\mu\nu\rho\sigma}$). In addition, we have that the trace of the $b_{2}$ term is zero on a Ricci-flat background for any operator. (This last fact is due to Ricci-flatness and complete tracelessness of non-vanishing $\hat{C}$ tensor.)

From~(\ref{formula42}) we might write
\begin{eqnarray}
\log Z^{2} &=& 3\log\det(\square_{1}/\mu^2)\nonumber \\[2mm]&& -2\log\det\left(\frac{\square_{2T}+2\hat{C}}{\mu^2}\right) \, .
\label{form440}
\end{eqnarray}
In the UV regime, when $\mu^{4}\gg C^2$, we can from \eqref{formula422} easily reproduce the flat spacetime tree-level result
\begin{eqnarray}
N_{{\rm DOF}}(C^2=0)&=&-(3\times4)+2\times9
\nonumber\\[2mm]
&=& -12+18 \ = \ 6\, .
\end{eqnarray}
%
%
Now the question stands, how does~(\ref{form440}) runs when changing the $C^2(x)$ or the $\mu$ scale of the RG energy? Take, for example, a determinant of the first vector term  
$\square_{1}/\mu^2$ in the numerator of~\eqref{formula42}, or a second one there in the denominator with the shifted argument $(\square_{2T}+2\hat{C})/\mu^2$ and in the traceless representation of the spin-2 fluctuation fields. To answer this, we use the same heat kernel methodology as in the previous subsection. Namely, by employing the fact that
\begin{eqnarray}
B'_{0}\left(\frac{\square_{2T}+2\hat{C}}{\mu^2}\right) \ = \ B'_{0}\left(\frac{\square_{2T}}{\mu^2}\right), 
\end{eqnarray}
and that
\begin{widetext}
\begin{eqnarray}
\log\det\left(\frac{\square_{2T}+2\hat{C}}{\mu^2}\right)
\ = -\frac{1}{(4\pi)^2}\left[\frac{1}{2}L^4B'_{0}\left(\frac{\square_{2T}}{\mu^2}\right) \ + L^2 B'_{2}\left(\frac{\square_{2T}+2\hat{C}}{\mu^2}\right) + \ B'_{4}\left(\frac{\square_{2T}+2\hat{C}}{\mu^2}\right)\log\frac{L^2}{\mu^2} \ + \ \ldots\right],~~~~
\end{eqnarray}
we get the Ricci-flat background analog of the expression~\eqref{ndof0}. In particular, 
for the contribution coming from the operator $\square_{2T}+2\hat{C}$ in \eqref{form440}, we get
\begin{eqnarray}
&&\mbox{\hspace{-3mm}}\frac{\frac{1}{2}L^{4}B'_{0}(\square_{2T}/\mu^2)+L^{2}B'_{2}\left(\frac{\square_{2T}+2\hat{C}}{\mu^2}\right)+B'_{4}\left(\frac{\square_{2T}+2\hat{C}}{\mu^2}\right)\log\frac{L^{2}}{\mu^{2}}}{\frac{1}{2}L^4B_{0}(\square_0/\mu^2)+L^{2}B_{2}(\square_0/\mu^2)+B{}_{4}(\square_0/\mu^2)\log\frac{L^{2}}{\mu^{2}}} \nonumber \\[2mm]
&&\mbox{\hspace{-3mm}}= \ \frac{\frac{1}{2}\frac{\mu^{4}}{\epsilon^{2}}B'_{0}(\square_{2T}/\mu^2)+\frac{\mu^{2}}{\epsilon}B'_{2}\left(\frac{\square_{2T}+2\hat{C}}{\mu^2}\right)- B'_{4}\left(\frac{\square_{2T}+2\hat{C}}{\mu^2}\right)\log\epsilon}{\frac{1}{2}\frac{\mu^{4}}{\epsilon^{2}}B_{0}(\square_0/\mu^2)+\frac{\mu^{2}}{\epsilon} B_{2}(\square_0/\mu^2)- B{}_{4}(\square_0/\mu^2)\log\epsilon}\nonumber \\[2mm]
&&\mbox{\hspace{-3mm}}= \ \frac{\frac{1}{2}B'_{0}(\square_{2T}/\mu^2)+\frac{\epsilon}{\mu^{2}}B'_{2}\left(\frac{\square_{2T}+2\hat{C}}{\mu^2}\right)-\frac{\epsilon^{2}}{\mu^{4}} B'_{4}\left(\frac{\square_{2T}+2\hat{C}}{\mu^2}\right)\log\epsilon}{\frac{1}{2}B_{0}(\square_0/\mu^2)+\frac{\epsilon}{\mu^{2}}B_{2}(\square_0/\mu^2)-\frac{\epsilon^{2}}{\mu^{4}} B{}_{4}(\square_0/\mu^2)\log\epsilon}\, .~~~~~~~
\label{III.B.36.yy}
\end{eqnarray}
\end{widetext}
We note again that this expression for the part of the ratio \eqref{formula422} is finite in both the UV and IR. Furthermore, we stress that for any Ricci-flat background we cannot factor out the integration over the 4-dimensional spacetime volume element, i.e. $\int d^4x\sqrt{|g|}$ [as we did in~(\ref{11.vb})], and formally cancel it out as a ``common factor'' between the numerator and the denominator in (\ref{formula42}). This is impossible, because on any Ricci-flat spacetime any scalar curvature invariant (such as the Kretschmann invariant is an example of) must have an explicit spacetime point dependence. 

In addition, the fact that the $b_{2}$ term is zero
on a Ricci-flat background, leads to a simplification of the previous expression to

\begin{widetext}
\begin{eqnarray}
(\ref{III.B.36.yy})&=&\frac{\frac{1}{2}B'_{0}(\square_{2T}/\mu^2)-\frac{\epsilon^{2}}{\mu^{4}} \ \!B'_{4}\left(\frac{\square_{2T}+2\hat{C}}{\mu^2}\right)\log\epsilon}{\frac{1}{2}B_{0}(\square_0/\mu^2)-\frac{\epsilon^{2}}{\mu^{4}} \ \! B{}_{4}(\square_0/\mu^2)\log\epsilon}\, .~~~~~
\end{eqnarray}
Consequently, taking sum of all two terms in (\ref{form440}) we can write 
\begin{eqnarray}
N_{\rm DOF}&=&-\frac{3\log\det(\square_{1}/\mu^2)-2\log\det\left(\frac{\square_{2T}+2\hat{C}}{\mu^2}\right)}{\log\det(\square_{0}/\mu^2)} \nonumber \\[2mm]
&=& \frac{6+6\frac{\epsilon^{2}}{\mu^{4}} B'_{4}\left(\square_{1}/\mu^2\right)\log\epsilon-4\frac{\epsilon^{2}}{\mu^{4}} B'_{4}\left(\frac{\square_{2}+2\hat{C}}{\mu^2}\right)\log\epsilon+4\frac{\epsilon^{2}}{\mu^{4}} B_{4}\left(\square_{0}/\mu^2\right)\log\epsilon}{1-2\frac{\epsilon^{2}}{\mu^{4}} B_{4}\left(\square_{0}/\mu^2\right)\log\epsilon}\, .
\end{eqnarray}
\end{widetext}
Using formulas (92) from the supplement of Ref.~\cite{RJ}, we have
\begin{eqnarray}
&&b_{4}\left(\square_{0}/\mu^2\right) \ = \ \frac{1}{180}C^{2} \ = \ \frac{1}{180}\mu^{4}c^{2}\, ,\nonumber \\[2mm]
&&
b'_{4}\left(\square_{1}/\mu^2\right)\ = \ -\frac{11}{180}C^{2} \ = -\frac{11}{180}\mu^{4}c^{2}\, ,\nonumber \\[2mm]
&&
b'_{4}\left(\frac{\square_{2T}+2\hat{C}}{\mu^2}\right) \ = \ \frac{189}{180}C^{2}\ = \ \frac{189}{180}\mu^{4}c^{2}\, ,\nonumber \\[2mm]
&&
b'_{4}\left(\frac{\square_{2}+2\hat{C}}{\mu^2}\right) \ = \ \frac{19}{18}C^{2}\ = \ \frac{19}{18}\mu^{4}c^{2}\, ,
\end{eqnarray}
where $c^{2} = C^2/\mu^4$ is the dimensionless square of the Weyl tensor characterizing
the background manifold. We might also note that for the Kretschmann invariant on the Ricci-flat background we have
\begin{eqnarray}
 K \ = \ C^2 \ = \ c^2\mu^4\, .
\end{eqnarray}
This relation is in analogy to the previous relations used on MSS backgrounds to define the dimensionless parameter $\lambda$ characterizing the curvature. Here this role is played by the dimensionless quantity $c^2$, which must be a spacetime dependent scalar field: $c^2=c^2(x)$. In the UV regime, when $\mu^{4}\gg C^2$, we have that $c^2(x)\ll1$.

By collecting our results, we find that the effective number of DOF is now given by
%
%
%
\begin{eqnarray}
&&\mbox{\hspace{-7mm}}N_{\rm DOF} \ = \ 
\frac{\int\! d^4x\sqrt{|g|}\left[6-\frac{137}{30}c^2(x)\epsilon^2\log\epsilon\right]}{\int\! d^4x\sqrt{|g|}\left[1-\frac{1}{90}c^2(x)\epsilon^2\log\epsilon\right]}\, .
\label{III.52.ccd}
\end{eqnarray}
In this formula, both in the numerator as well as in the denominator, we do integrate over the spacetime volume manifold of a Ricci-flat background, independently in the numerator and in the denominator. See Appendix~\ref{Appendix C} for more details of the above calculation of the
the ratio of the scalar heat kernel coefficients in the Ricci-flat background case. 
In particular, in Appendix~\ref{Appendix C}, we explained how one can go from the ratio of spacetime-integrated heat-kernel coefficients ($B$'s) to the ratio of the corresponding integral densities, where we use only un-integrated coefficients, i.e. $b$'s.

As a result, we see that it is sufficient to plot the following function 
\begin{eqnarray}
N_{\rm DOF} \ = \ \frac{6-\frac{137}{30}\alpha\epsilon^2\log\epsilon}{1-\frac{1}{90}\alpha\epsilon^2\log\epsilon}\, ,
\label{form26}
\end{eqnarray}
where
\begin{eqnarray}
\alpha \ = \ \int\! d^4x\sqrt{|g|}\ \!  c^2(x)\, ,
\end{eqnarray}
is a constant parameter that depends on the actual Ricci-flat background chosen. 
The plots of this running for various values of the parameter $\alpha$ are depicted in Fig.~\ref{Fig.2}, where we took into account only the numerator of the formula \eqref{form26}. Now, we see that the effective number of degrees of freedom is only a dimensionless function of a dimensionless UV-regulator $\epsilon$  and of dimensionless parameter $\alpha$ characterizing the Ricci-flat backgrounds.
\begin{figure}[!ht]
\begin{center}
$\mbox{\hspace{4mm}}$\includegraphics[width= 8.0cm]{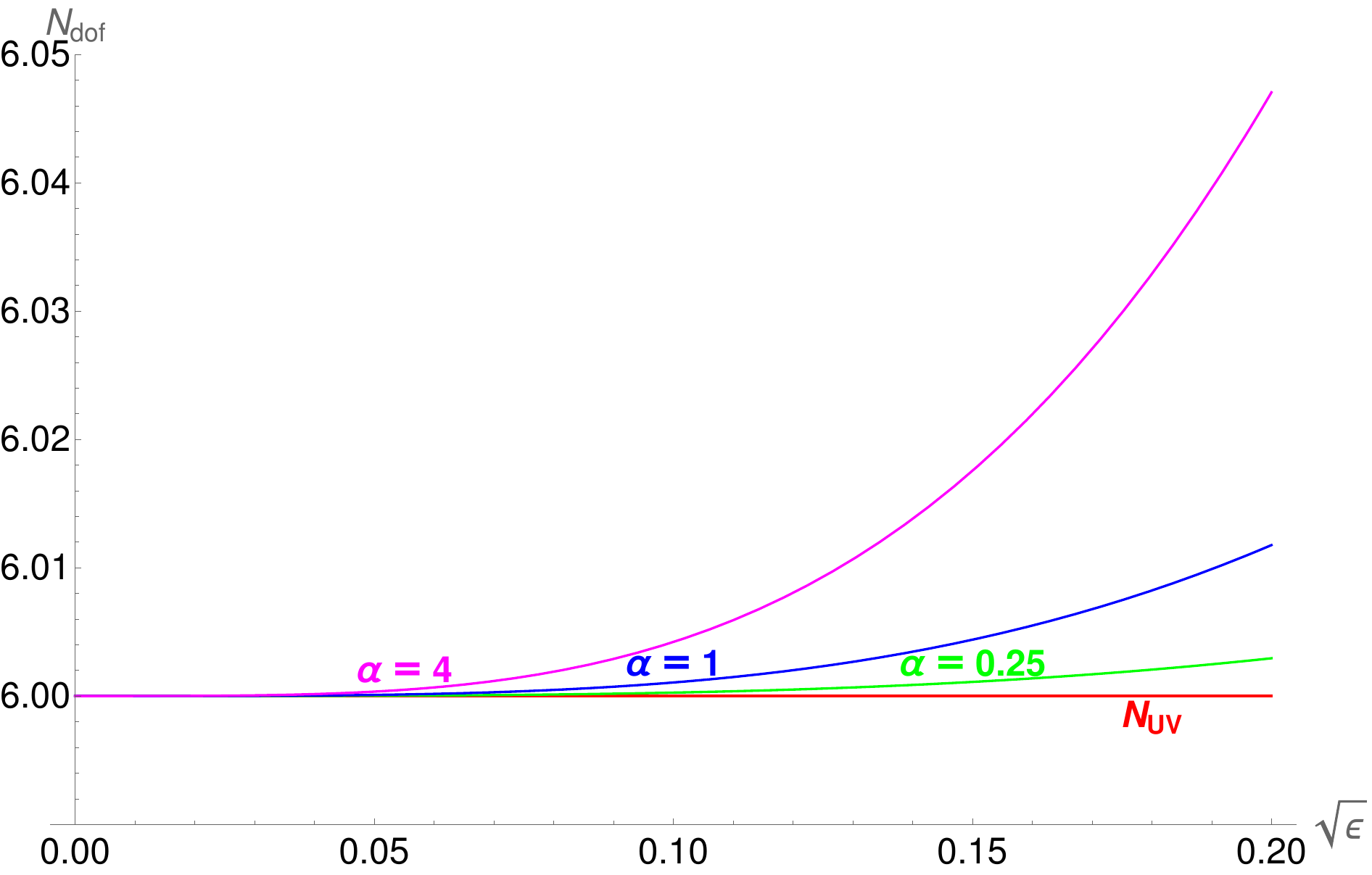}
\vspace{1mm}
\caption{Plot of $N_{\rm DOF}$ as a function of $\sqrt{\epsilon}=\mu / L$ for various positive values of the dimensionless parameter $\alpha$. In red we show the value of DOF in the UV limit. For simplicity, only the numerator of \eqref{form26} is considered here.}
\label{Fig.2}
\end{center}
\end{figure}
We see that at the limiting point $\epsilon=0$ the number of degrees of freedom reaches six as it should for the conformal quantum gravity in the UV. 

By taking in~(\ref{form26}) the derivative with respect to $\epsilon$ at the point $\epsilon=1/25$ and keeping $\alpha$ fixed, one obtains that
%
%
%
%
%
\begin{eqnarray}\label{form44}
\left.\frac{d N_{\rm DOF}}{d\epsilon}\right|_{\epsilon=1/25} \!\! =  \frac{569531250 \ \!\alpha \ \! (\log 625 - 1)}{(56250 + \alpha \log25)^2}  > 0 \, .
\end{eqnarray}
Here we assumed that the possible values of the $\alpha$ parameter are all positive as this is related to the integrated Kretschmann invariant of the Euclidean manifold.
Thus for all values of $\alpha$, DOF increase (rather than decrease)
toward the IR regime. This is certainly not desirable. Since this behavior is true for any value of $\alpha$, it cannot be ascribed to our perturbation treatment.

From formulas (\ref{form20}) and (\ref{form44}), and also by comparing Fig.~\ref{Fig.1}. and Fig.~\ref{Fig.2}., we can clearly see that the running of DOF on different backgrounds behave in a qualitatively different behavior. This means that the results obtained by the FT method, computed via the background field method, in the one-loop order are strongly dependent on the chosen (Bach-flat) background and is therefore not suitable in our case. In addition, we could see that there exists a large parameter space where the FT method does not provide the expected reduction in DOF. Therefore, in order to obtain meaningful results independent of the background, we need to use a different technique. To this end, we now turn to the method based on trace anomaly and ensuing the $c$- and $a$-function.

 \section{Flow using beta function $\beta_C$ and $c-$anomaly in Quantum Weyl Gravity\label{Sec.III}}

 {Since we want to associate the number of degrees of freedom with the start and end points of RG flows in a universal way, it is extremely important to identify quantities that are invariant or exhibit a regular, predictable behavior along RG flows. One celebrated  example is due to 't Hooft~\cite{tHooft:1979rat}, who identified  't Hooft anomalies of global symmetries as RG-independent quantities, which therefore determine the so-called ``anomaly matching conditions''  that constrain the possible RG flows. On the other hand, trace anomalies are related to conformal symmetry, which is broken by a generic RG flow, so that any kind of anomaly matching seems to be meaningless in this context. Yet, it is still possible to identify protected quantities that are universal in the sense that they do not depend on the details of the RG flow~\cite{Komargodski:2011vj,Komargodski:2011xv,key-31,key-32,Luty:2012ww,Baume:2014rla}. In $d=2$ and $d=4$ RG flows, for example, it has been noticed that, if the QFT is coupled to a massless dilaton, at low energies the dilaton-dilaton scattering amplitude is universally proportional to $\Delta c = c_{\rm{UV}}-c_{\rm{IR}}$ and $\Delta a = a_{\rm{UV}}-a_{\rm{IR}}$, respectively, where $a$ and $c$ are the coefficients of the corresponding terms in the trace anomalies of the UV and IR CFTs. This remarkable fact has been used to rederive the classical result of Zamolodchikov in $d=2$ as well as to put on a firmer ground an analogous expected result in $d=4$. $\Delta c$ in $d=2$ QFTs has a very physically suggestive interpretation due to Cardy's fundamental result that the asymptotic density of states of a unitary CFT is universally given in terms of its central charge~\cite{Cardy:1986ie}. This allows to determine that for CFTs with $c>1$ the spectrum must include an infinite number of Virasoro primaries and that every integer spin must appear in a bosonic 
 CFT~\cite{Mukhametzhanov:2019pzy,Mukhametzhanov:2020swe}. A similar interpretation for $\Delta a$ in $d=4$ QFTs is unfortunately lacking.  Recently, however, the quantity $\Delta c-\Delta a$ has been universally connected to the helicity flipping term in the graviton-dilaton amplitude at low energy providing a relation to spinning massive states in spectrum of the QFT coupled to the dilaton and graviton background~\cite{Karateev:2023mrb}. On the other hand, the combination $c-a$ has been seen to generalize Cardy formula to $d=4$ in the context of counting specific operators in supersymmetric theories through the so-called superconformal index~\cite{DiPietro:2014bca,Beem:2017ooy}. Moreover $c-a$ was related to spinning primary operators in the large central charge, strong coupling limit of CFTs~\cite{Camanho:2014apa}.} Connection of $a$- and $c$-function with the number of DOF in $d=4$ was firstly studied in~\cite{duff}. The fact that the trace anomaly can be understood as a measure of DOF was first discussed in~\cite{Cappelli:00}. In the following we will take this as our ``modus operandi'' and explore ensuing consequences for QCG.


We now utilize the fact that the $c$-anomaly  is proportional to the beta function $\beta_C$ (at least to two loops). The $c$-anomaly is part of the trace anomaly directly proportional to the term $C^2$ of the effective action. This can be clearly seen from the trace anomaly formula~\cite{duff}
\begin{eqnarray}
\langle T^{\mu}_{~\mu} \rangle &=& 
\frac{2}{\sqrt{g}} g^{\mu \nu} \frac{\delta \Gamma}{\delta  g^{\mu \nu}}
\ = \ c \,C^2 \ + \ \cdots \nonumber \\[2mm] &=&  \frac{\beta_C }{(4\pi)^2} \ \! C^2 \ + \  \cdots \,.
\label{IV.57.cv}
\end{eqnarray}
Therefore when we normalize properly the one-loop and RG-improved expression for the $\beta$ function $\beta_C$, we get another form of the running of the effective number of degrees of freedom in Weyl gravity. The aforementioned RG improvement takes into account threshold phenomena and the effect of the anomalous dimension of the graviton operator. In~\cite{RJ}, it was found that the RG-improved $\beta_C$ satisfies
 \begin{equation}
     \beta_C \ = \frac{1}{4\pi^2} \frac{b-X}{1+y(X-b)}\, ,\label{form28}
 \end{equation}
with the definition of constants $b=137/60$ and $y=2/\omega_C$, where $\omega_C/2$  is the coupling in front of the Weyl term, i.e. $\propto 1/\alpha_{\rm{w}}^2$. For simplicity, one can take $y=0$ (or effectively when the coupling $\omega_C$ is very big:  $\omega_C\gg1$) and neglect here the sub-leading effect of the anomalous dimension. The expression for $X$ reads explicitly~\cite{RJ}
\begin{eqnarray}
\mbox{\hspace{-5mm}}X &=& - \ \frac{21}{40}\left(1-\frac{\frac{2}{3}\Lambda}{\nu^{2}}\right)^{-1} \ + \ \frac{9}{40}
\left(1-\frac{\frac{4}{3}\Lambda}{\nu^{2}}\right)^{-1}\nonumber \\[2mm]
&-&  \frac{179}{45}
\left(1+\frac{\Lambda}{\nu^{2}}\right)^{-1} \ - \ \frac{59}{90}\left(1+\frac{\frac{1}{3}\Lambda}{\nu^{2}}\right)^{-1}\nonumber \\[2mm]
&+ & \frac{479}{360}\left(1+\frac{2\Lambda}{\nu^{2}}\right)^{-1} \ - \
\frac{269}{360}\left(1+\frac{\frac{4}{3}\Lambda}{\nu^{2}}\right)^{-1}.
\label{Xeqn2}
\end{eqnarray}
In this and the following sections, we decide to call the energy scale $\nu$ in order to distinguish it from the sliding scale $\mu$ of the previous section. Later we can compare between the two. We only note here that the limits for the variable $\nu$ are opposite to those used in the previous section for the sliding scale $\mu$, namely $\nu\to+\infty$ corresponds to the UV regime, while $\nu\to0$ gives us an IR regime. 
When we use the proper normalization (saying that we get precisely six degrees of freedom in the UV regime), we get the following formula for the effective running number of degrees of freedom,
 \begin{equation}
     N_{\rm DOF} \ = \ \frac{6}{\frac{199} {30}} {4\pi^2}\beta_C \ = \ \frac{180}{199} {4\pi^2}\beta_C\,  .
 \end{equation}
One sees that this formula also describes the monotonically decreasing number of DOF, which starts with 6 DOF in the UV. We shall not trust this running after the tentative IR FP, where $\beta_C=0$ since this would imply a non-positive number. The coefficient $\frac{180}{199}$ is numerically quite close to unity, so the plot gives a qualitatively sensible result for the effective number of degrees of freedom in QCG.

When one takes the logarithmic derivative of the flow of the number of DOF (from the numerator of the formula (\ref{form28}) only), that is $\nu \frac{dN_{\rm DOF}}{d\nu}$ evaluated at the reference point with $\nu=L/5$, one obtains 
\begin{eqnarray}
&&\left.\nu \frac{d N_{\rm DOF}}{d\nu}\right|_{\nu=\frac{1}{5}L} \ = \left. -\frac{180}{199}\nu \frac{d X}{d\nu}\right|_{\nu=\frac{1}{5}L}\nonumber\\[2mm]
&&= -\frac{50}{597}\frac{\Lambda}{L^2}  \left(\frac{189}{\left(1-\frac{50 \Lambda }{3
   L^2}\right)^2}-\frac{118}{\left(1+\frac{25\Lambda }{3
   L^2}\right)^2}-\frac{2148}{\left(1+\frac{25\Lambda
   }{L^2}\right)^2}\right.\nonumber\\[2mm]
   &&\;\;\;\;- \left.\frac{538}{\left(1+\frac{100 \Lambda }{3
   L^2}\right)^2}+\frac{1437}{\left(1+\frac{50 \Lambda
   }{L^2}\right)^2}-\frac{162}{\left(1-\frac{100 \Lambda }{3
   L^2}\right)^2}\right)\nonumber\\[2mm]
   &&= \ \frac{50}{199}\left(\frac{1340\, \Lambda }{3 L^2}+\frac{8900\, \Lambda ^2}{9
   L^4}+O\left(\Lambda ^3\right)\right)\nonumber\\[2mm]
   &&= \ \frac{67\,000\, \Lambda }{597 L^2} \ + \ \frac{445\,000\, \Lambda ^2}{1791
   L^4}\ + \ \mathcal{O}\left(\Lambda ^3\right) \ > \ 0\,,
   \label{derivative1}
\end{eqnarray}
which is qualitatively consistent with the quoted results on MSS in (\ref{form13}).

Two final comments are in order here. Firstly, we read the $\beta$ function $\beta_C$ from the combination $\beta_C+\beta_E$ which is evaluated on the Ricci-flat backgrounds. The Weyl tensor vanishes on MSS background, hence just from there the $\beta$ function $\beta_C$ cannot be obtained and disentangled. Second, for non-trivial running ($\nu$-dependence) in the expression $X$ one must have a non-vanishing $\Lambda$ parameter of the MSS background, so the use of these two backgrounds is necessary here.

\section{Flow using the beta function $\beta_E$ (or $a-$anomaly) in Quantum Weyl Gravity\label{Ref.IV}}

The $\beta$ function $\beta_E$  is proportional to the $a$-anomaly and it can be used to directly count the number of degrees of freedom via the celebrated 4-dimensional $a$-theorem. The $a$-anomaly is part of the trace anomaly directly proportional to the Gauss--Bonnet term  (also known as the Euler topological term) $E_4$  of the effective action. One can write the trace anomaly in the form
\begin{equation}
     \langle T^{\mu}_{~\mu} \rangle \ = \  a\, E_4 \ + \  \cdots  \ = \ \frac{\beta_E }{(4\pi)^2} \ \! E_4 \ + \  \cdots \,.
     \label{V.62.cb}
 \end{equation}
 Similarly as in the previous section, when we normalize properly the one-loop and RG-improved expression for the $\beta$ function $\beta_E$, we will get another schematic form of the running of the effective number of DOF in Weyl gravity. Again, the improvement takes into account threshold phenomena and the effect of the anomalous dimension of the graviton operator. In~\cite{RJ}, it was found that $\beta_E$ satisfies
 \begin{equation}
     \beta_E \ = \  \frac{1}{4\pi^2} \frac{X}{1+y(X-b)}\, ,\label{form282}
 \end{equation}
with the definition of constants $b=137/60$ and $y=2/\omega_C$, where $\omega_C \propto 1/\alpha_{\rm w}^2$. For simplicity, one can take $y=0$ (or effectively when the coupling $\omega_C$ is very big, i.e. $\omega_C\gg1$) and neglect here the sub-leading effect of the anomalous dimension. The explicit expression $X$ reads 
\begin{eqnarray}
\mbox{\hspace{-5mm}}X &=& - \ \frac{21}{40}\left(1-\frac{\frac{2}{3}\Lambda}{\nu^{2}}\right)^{-1} \ + \ \frac{9}{40}
\left(1-\frac{\frac{4}{3}\Lambda}{\nu^{2}}\right)^{-1}\nonumber \\[1mm]
&-&  \frac{179}{45}
\left(1+\frac{\Lambda}{\nu^{2}}\right)^{-1} \ - \ \frac{59}{90}\left(1+\frac{\frac{1}{3}\Lambda}{\nu^{2}}\right)^{-1}\nonumber \\[1mm]
&+ & \frac{479}{360}\left(1+\frac{2\Lambda}{\nu^{2}}\right)^{-1} \ - \
\frac{269}{360}\left(1+\frac{\frac{4}{3}\Lambda}{\nu^{2}}\right)^{-1}\,.
\label{Xeqn3}
\end{eqnarray}
We remark that the value of  $\beta_E$  in the UV regime of QCG, i.e 
\begin{equation}
    \beta_E^{\rm UV} \ = -\frac{1}{4\pi^2} \frac{87}{20} \ < \ 0\, ,
\end{equation}
was already found in~\cite{Fradkin:1985am}.
When we use the proper normalization (saying that we get precisely six DOF in the UV regime), we get the following formula for the effective running number of DOF,
 \begin{equation}
     N_{\rm DOF} \ = \  -\frac{6}{\frac{87}{20}} {4\pi^2}\beta_E \ = \ -\frac{40}{29}{4\pi^2}\beta_E\, .
 \end{equation}
It can be seen that this formula also describes the monotonically decreasing number of DOFs, starting with 6 DOFs in the UV regime. We shall not trust this running after the tentative IR FP, where $\beta_E=0$ since this would imply a non-positive number. The coefficient $\frac{87}{20}$ is numerically quite close to four, so the plot in Fig~\ref{Fig.3} gives a qualitatively sensible result for the effective number of degrees of freedom in QCG.

When one takes the logarithmic derivative of $N_{\rm DOF}$ (from the numerator of the formula (\ref{form282}) only), that is $\nu \frac{dN_{\rm DOF}}{d\nu}$ evaluated at the refrence point with $\nu=L/5$, one obtains
\begin{eqnarray}
&&\left.\nu \frac{d N_{\rm DOF}}{d\nu}\right|_{\nu=\frac{1}{5}L} \ = \left. -\frac{40}{29}\nu \frac{d X}{d\nu}\right|_{\nu=\frac{1}{5}L}\nonumber\\[2mm]
&&= -\frac{100}{783}\frac{\Lambda}{L^2}  \left(\frac{189}{\left(1-\frac{50 \Lambda }{3
   L^2}\right)^2}-\frac{118}{\left(1+\frac{25\Lambda }{3
   L^2}\right)^2}-\frac{2148}{\left(1+\frac{25\Lambda
   }{L^2}\right)^2}\right.\nonumber\\[2mm]
   &&\mbox{\hspace{3mm}}- \left.\frac{538}{\left(1+\frac{100 \Lambda }{3
   L^2}\right)^2}+\frac{1437}{\left(1+\frac{50 \Lambda
   }{L^2}\right)^2}-\frac{162}{\left(1-\frac{100 \Lambda }{3
   L^2}\right)^2}\right)\nonumber\\[2mm]
   && = \ \frac{134\,000\, \Lambda }{783\,L^2} \ + \ \frac{890\,000\, \Lambda ^2}{2349\,
   L^4}\ + \ \mathcal{O}\left(\Lambda ^3\right) \ > \ 0\,,\label{derivative2}
\end{eqnarray}
which is consistent with the quoted results on MSS in (\ref{form13}) and qualitatively this is the same order of magnitude of values as in the previous formula in (\ref{derivative1}).

The plots of the two running functions describing the effective number of degrees of freedom in QCG proportional respectively to $\beta_C$ and $\beta_E$ beta functions versus energy scale $\nu$ measured in units of $\sqrt{\Lambda}$ are depicted in Fig.~\ref{Fig.3}. Actually, the two derivatives in formulas \eqref{derivative1} and \eqref{derivative2} are both positive and  proportional to each other. 
\begin{figure}[!ht]
\begin{center}
$\mbox{\hspace{4mm}}$\includegraphics[width= 8.0cm]{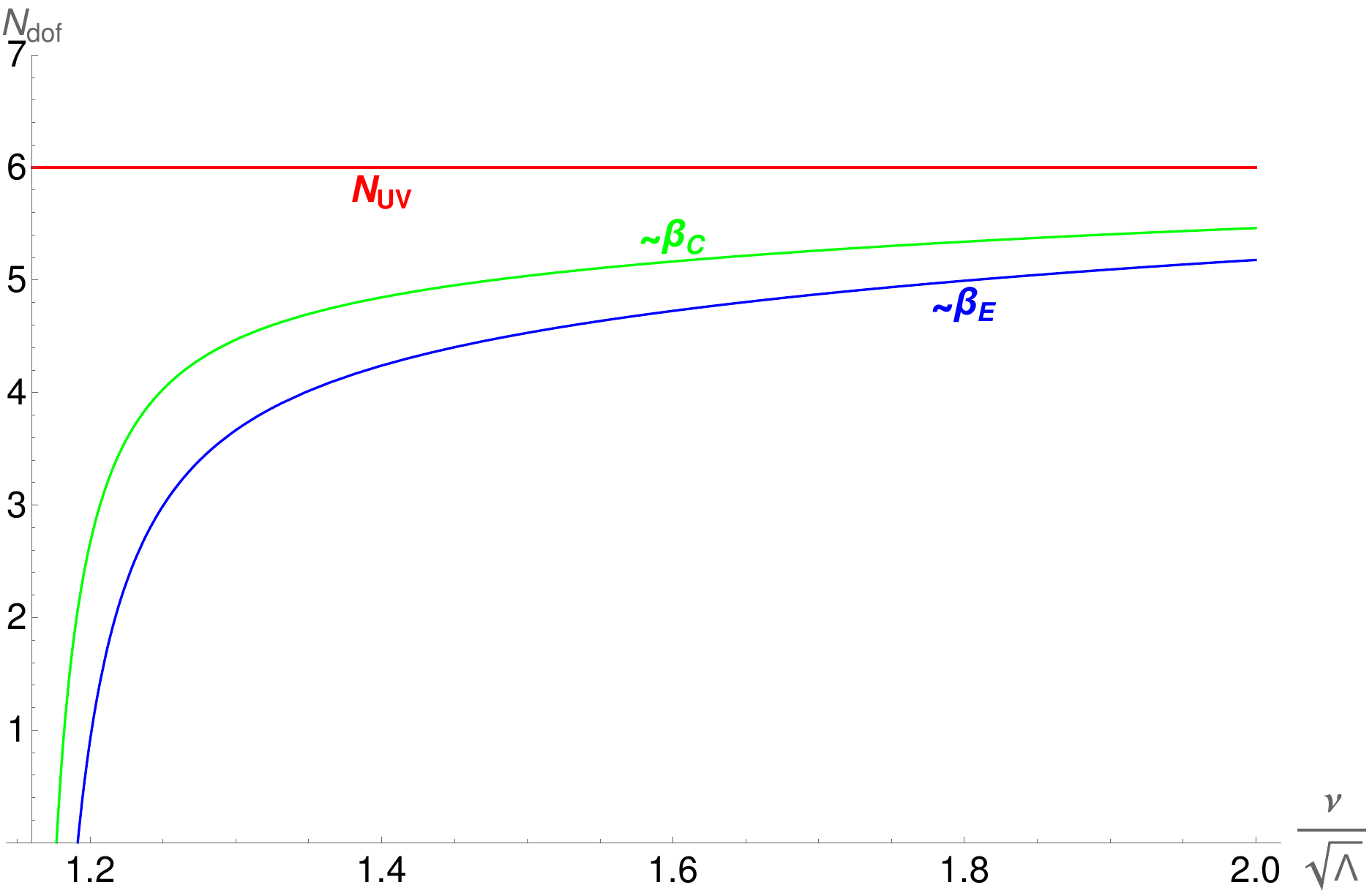}
\vspace{1mm}
\caption{Plot of $N_{\rm DOF}$ as a function of $\nu / \sqrt{\Lambda}$ for the calculations based on the two anomalies (proportional to $\beta_C$ and $\beta_E$ respectively). In red we show the value of DOF in the UV limit.}
\label{Fig.3}
\end{center}
\end{figure}

\section{Connection between  FT method and  trace-anomaly approach\label{Sec.V}}

As we have seen in Section~\ref{Sec.II}, the FT method yields background dependent result for DOF. On the other hand, the  method based on the trace anomaly is background independent.  Moreover, the coefficients $a$ and $c$ cannot be obtained as the variation of a local functional and are therefore not affected by the renormalization ambiguities. In last two sections 
we computed the $a$- and $c$-anomaly 
by employing the identity
\begin{eqnarray}
  \langle T^{\mu}_{~\mu} \rangle &=& 
  a\ \! E_4 \ + \ c\ \!C^2  \ + \  \cdots \nonumber \\[2mm]  
  &\propto& \beta_E\ \! E_4 \ + \  \beta_C \ \! C^2 \ + \ \cdots \, .
\end{eqnarray}
Here the second line becomes identity when we multiply it by the factor $1/(4\pi)^2$, see Eqs.~(\ref{IV.57.cv}) and~(\ref{V.62.cb}). Although the second
identity is formally valid only up to two loops due to the renormalization-prescription dependence of the $\beta$ functions~\cite{JRGK}, it is very useful because it allows to compute the running of the DOF directly from the $\beta$ functions --- fact that we utilized in previous two sections.

Despite the different methodologies, it is interesting to note that both of the above approaches for DOF computation are closely related. To find the connection, we employ two relations. First, we consider a rescaling of the background metric by {\em constant} factor $\Omega^2$, i.e. $g^{(0)}_{\mu\nu} \mapsto \tilde{g}^{(0)}_{\mu\nu} = \Omega^2 g^{(0)}_{\mu\nu}$.  In this case the effective action transforms as 
\begin{eqnarray}
\Gamma[\Omega^2 g^{(0)}] \ = \ \Gamma[g^{(0)}] \ - \ 2 (\log \Omega) \ \! \zeta_{\Delta}(0)\, ,
\label{Sec.V.57.nn}
\end{eqnarray}
where $\zeta_{\Delta}(0)$ is a zeta function of the kernel operator from the classical action.
A derivation of this identity is provided in Appendix~\ref{Appendix B}. Eq.~(\ref{Sec.V.57.nn}) directly implies that
\begin{eqnarray}
\left. \frac{d\Gamma[\tilde{g}^{(0)}]}{d \Omega^2}\right|_{\Omega = 1 } \ = \  \ - \  \zeta_{\Delta}(0)\, .
\label{Sec.V.58.cv}
\end{eqnarray}
On the other hand, if we consider an infinitesimal rescaling $\Omega = 1 + \omega  \approx 1$, we might write
\begin{eqnarray}
\delta_{\omega} \Gamma &=& \int d^4 x \ \! \left[ \frac{\delta \Gamma}{ \delta h^{\mu \nu }} (-2\omega) h^{\mu\nu} \ + \  \frac{\delta \Gamma}{ \delta {\mathcal{C}}_{\mu}} d_{\mathcal{C}} \omega \ \! {\mathcal{C}}_{\mu}  \right. \nonumber\\[2mm]
&& \left. + \ \frac{\delta \Gamma}{ \delta {\bar{\mathcal{C}}}^{\mu}} d_{\bar{\mathcal{C}}} \omega \ \!{\bar{\mathcal{C}}}^{\mu} \
+ \ \frac{\delta \Gamma}{\delta g^{(0)\mu \nu}} (-2\omega)  g^{(0)\mu \nu} \right] \nonumber \\[2mm]
&=& - \omega \int d^4 x \ \! \sqrt{g} \ \! g^{(0)\mu \nu} \langle T_{\mu \nu} \rangle\, , 
\label{V.59.cc}
\end{eqnarray}
where on the last line we have used the equations of motion for the metric fluctuation field $h^{\mu\nu}$ and the Faddeev--Popov fields ${\bar{\mathcal{C}}}^{\mu}$ and ${{\mathcal{C}}}_{\mu}$. Coefficients $d_{\bar{\mathcal{C}}} $ and $d_{\mathcal{C}}$ represent  scale dimensions of the fields ${\bar{\mathcal{C}}}^{\mu}$ and ${{\mathcal{C}}}_{\mu}$, respectively.
The scalar term $\sqrt{g}$ is also computed with respect to the background metric.

By comparing (\ref{V.59.cc})
with (\ref{Sec.V.57.nn}), we obtain
\begin{eqnarray}
   \int  {d}^4x \ \! \sqrt{g} \ \!  \langle T^{\mu}_{~\mu} \rangle  \ = \ 2 \zeta_{\Delta}(0)\, ,
\end{eqnarray}
and by employing Eq.~(\ref{Sec.V.58.cv}), we can rewrite the integrated trace anomaly in the form
\begin{eqnarray}
\int  {d}^4x \ \! \sqrt{g} \ \!  \langle T^{\mu}_{~\mu} \rangle  &=& - 2\left.\frac{d\Gamma[\tilde{g}^{(0)}]}{d \Omega^2}\right|_{\Omega = 1 } \nonumber \\[2mm] &=& \left. \dfrac{{d} \log Z^2}{{d}\Omega^2}\right|_{\Omega = 1 }\, .
\label{hawking-scaling}
\end{eqnarray}
%
%
%
%
Eq.~(\ref{hawking-scaling}) is the sought relation.

Several comments are in order at this point.
First, the FT method is primarily based on the computation of $\log Z^2$, while the $a$- and $c$-anomaly approaches are based on the trace anomaly (and, in particular, $\beta$-function evaluation). As these are conceptually distinct types of computations, the formula (\ref{hawking-scaling}) allows to
check the mathematical consistency of both methods. Second, since MSS are conformally flat spaces, we can plug the expression for the trace into the formula (\ref{hawking-scaling}) and get 
\begin{equation}
    32\pi^2 a \ \!\chi \ = \ 2 \beta_E \ \!\chi \ = \  \left. \dfrac{{d} \log Z^2}{{d}\Omega^2}\right|_{\Omega = 1 }\, ,
    \label{71.cc}
\end{equation}
where $\chi$ is the Euler--Poincaré invariant (\ref{II.8.aa}) (also known as the full spacetime volume integral of the Gauss--Bonnet density term).
%
%
So, we see that the partition function on conformally flat spaces selects only the $a$ anomaly out of the full trace anomaly. In passing we note that $\chi$  on the $d$-dimensional sphere $S^d$ is two, i.e.  $\chi(S^d) = 2$.
%
%

The argument leading to (\ref{71.cc}) cannot be applied in the case of e.g. Ricci-flat backgrounds, since then the integral on the left-hand side of (\ref{hawking-scaling}) cannot be evaluated in general to obtain a simple relation between $\log Z$ and $a$- or $c$-anomaly. The need to select a specific background once again illustrates the background dependence of the FT method.

The usefulness of Eq.~(\ref{71.cc}) can be illustrated with the number of DOF in the UV regime. In such a case, the $\Lambda$ terms in (\ref{formula22}) can be neglected compared to the $\Box$ contribution, and the one-loop partition function can be written as
\begin{eqnarray}
Z_{{\rm 1-loop}}^{\rm{uv}} \ = \ \left( {\rm{det}}_{0} \Box \right)^{-N_{\rm{DOF}}^{\rm{uv}}/2} \, ,
\end{eqnarray}
and hence
\begin{eqnarray}
\log Z_{{\rm 1-loop}}^{\rm{uv}} &=&- \frac{N_{\rm{DOF}}^{\rm{uv}}}{2}  \ \!{\rm{Tr}} \log \left(\Box_0/\mu^2\right) \nonumber \\[2mm]
&=&  - \frac{N_{\rm{DOF}}^{\rm{uv}}}{2}  \ \! \sum_{n} \log \omega_n\, ,
\end{eqnarray}
where  ${\omega}_n$ are the (dimensionless) eigenvalues of $\Box_0/\mu^2$. Following the argument in Appendix~\ref{Appendix B}, we write
\begin{eqnarray}
&&\mbox{\hspace{-10mm}}\log Z_{{\rm 1-loop}}^{\rm{uv}}(\Omega^2 g^{(0)}) \nonumber \\[2mm]&=& - \frac{N_{\rm{DOF}}^{\rm{uv}}}{2}  \ \! \sum_{n} \log \left( \Omega^{-2}\omega_n\right)\nonumber \\[2mm]
&=& - \frac{N_{\rm{DOF}}^{\rm{uv}}}{2}  \ \! \sum_{n} \log \left( \omega_n\right) \ + \ N_{\rm{DOF}}^{\rm{uv}} \log \Omega  \ \! \zeta_{\Box}(0)\nonumber \\[2mm]
&=& \log Z_{{\rm 1-loop}}^{\rm{uv}}(g^{(0)}) \ + \ N_{\rm{DOF}}^{\rm{uv}} \log \Omega  \ \! \zeta_{\Box}(0) \, .~~~~
\label{VI.77.cv}
\end{eqnarray}
The spectral $\zeta$-function $\zeta_{\Box}(0)$ is finite when analytically continued from $\rm{Re}(s) > 2$. We also took advantage of the fact that in the UV regime $N_{\rm{DOF}}^{\rm{uv}}$ is a background (and thus $\Omega$) independent constant, see Fig.~\ref{Fig.1}.  
Relations~(\ref{71.cc}) and~(\ref{VI.77.cv}) allow to write 
\begin{eqnarray}
32\pi^2 a^{\rm{uv}} \ \!\chi  &=&  N_{\rm{DOF}}^{\rm{uv}} \ \! \zeta_{\Box}(0)  \nonumber \\[2mm] &=&  N_{\rm{DOF}}^{\rm{uv}} \ \! \left(32\pi^2 a^{\rm{uv}}_1 \ \!\chi \right),~~
\end{eqnarray}
where $a^{\rm{uv}}_1$ is the $a$-function corresponding to a massless scalar field in the UV. Consequently
\begin{eqnarray}
N_{\rm{DOF}}^{\rm{uv}} \ = \ \frac{a^{\rm{uv}}}{a^{\rm{uv}}_1}\, ,
\end{eqnarray}
as could be expected.

\section{Conclusions \label{Concl.}}

Quantum conformal gravity provides interesting, conceptually rich, and largely uncharted territory among higher-derivative gravity theories.
Motivated by the recent suggestion that QCG, after (global) conformal symmetry breaking, approaches the Starobinsky $f(R)$ gravity in its low energy phase~\cite{RJ, JKS, JRGK}, we have analyzed in this paper how the number of DOF in QCG varies 
when moving away from the UV fixed point. 
This study is part of a larger research project whose objective is to elucidate the manner in which the Weyl symmetry breaks down in QCG. 
To this end, we used the conventional background field method and employed two methods of counting: the Fradkin--Tseytlin partition function-based prescription and the method based on the trace anomaly and ensuing $c$- and $a$-function.

In order to implement the FT prescription, we have computed partition functions involved in the one-loop approximation and for two Bach-flat spacetimes, namely MSS and Ricci-flat background. Our computations relied on the one-loop results from Ref.~\cite{JKS} where the York decomposition of the metric field  was employed.  It turned out that the resulting counting of DOF is very sensitive to the choice of background, and in particular it exhibits qualitatively and quantitatively different running on MSS and Ricci-flat background.  This indicates that the FT method does not seem to be suitable for calculating the running of DOF --- at least not to one-loop accuracy. 
On the other hand, the trace anomaly-based method produced consistent results for both aforementioned backgrounds with the expected decrease in DOF as the energy scale lowered from UV to IR. 
To compute the $a$- and $c$-function, we used the fact that they are proportional to the corresponding $\beta$ functions (at least at the two-loop level). 
Our computations of $\beta$ functions relied on the enhanced one-loop scheme~\cite{JKS}, where the enhancement refers to two non-perturbative effects that were included, namely 1) threshold phenomena and 2) the effect of the anomalous dimension.

Finally, we have shown how the trace of the energy
momentum is related to the behaviour of the partition function under scale transformations. This allowed us to establish a link between the two discussed methods and to confirm the expected behavior in the UV regime.




\begin{acknowledgments}
P.J. acknowledges the support from the Czech Ministry of Education grant M\v{S}MT RVO 14000.
J.K. was supported by the Grant Agency of the Czech Technical University in Prague, grant No. SGS22/178/OHK4/3T/14.
The research of S.G.\ has been supported by a BIRD-2021 project (PRD-2021) and by the  PRIN Project n.~2022ABPBEY, ``Understanding quantum field theory through its deformations''.
We also wish to thank Roberto Percacci for useful discussion, comments and suggestions in the initial stage of the project resulting in this article. 
\end{acknowledgments}


\vspace{5mm}
\appendix

\section{Toy computation in Einsteinian  gravity in $d=4$ \label{Appendix A}}

In this appendix we use Einstein gravity (EG) as a test bed  for the FT prescription for
DOF. The computation is again done for two backgrounds --- 4-dimensional MSS background and a Ricci-flat background. 
Besides being Bach-flat, these backgrounds are also 
Einstein spaces, which are characterized by the condition that $R_{\mu \nu } = \Lambda g_{\mu\nu}$ with
$\Lambda =$~const. Einstein spaces are 
vacuum solutions of Einstein's equation with $\Lambda$ proportional to the cosmological constant.

\subsection{MSS background}

Properties of a 4-dimensional MSS manifold are specified at the beginning of Section~\ref{Sec.IIIA}. The square of the partition function of EG at the one-loop level reads
\begin{equation}
Z^{2} \ = \ \frac{\det^{2}\left(\frac{\square_{1}+\Lambda}{\mu^{2}}\right)}{\det\left(\frac{\square_{2T}-\frac{2}{3}\Lambda}{\mu^{2}}\right)\det\left(\frac{\square_{0}+2\Lambda}{\mu^{2}}\right)}\, .
\end{equation}\\
The effective number of DOF then reads
\begin{eqnarray}
N_{{\rm DOF}} &=& -\frac{2\log\det\left(\frac{\square_{1}+\Lambda}{\mu^2}\right)}{\log\det\left(\frac{\square_{0}}{\mu^2}\right)} \nonumber \\[2mm] &+& 
\frac{\log\det\left(\frac{\square_{2T}-\frac{2}{3}\Lambda}{\mu^2}\right)+\log\det\left(\frac{\square_{0}+2\Lambda}{\mu^2}\right)}{\log\det\left(\frac{\square_{0}}{\mu^2}\right)}\, ,~~~~
\end{eqnarray}
which, in terms of the heat kernel coefficients, has the explicit form  
\begin{widetext}
\begin{eqnarray}
N_{{\rm DOF}}  &=& -\ \frac{\frac{1}{2}L^{4}\left(2b'_{0}\left(\frac{\square_{1}+\Lambda}{\mu^2}\right)\right)}{\frac{1}{2}L^{4}b_{0}\left(\frac{\square_{0}}{\mu^2}\right)+L^{2}b_{2}\left(\frac{\square_{0}}{\mu^2}\right)+\log\frac{L^{2}}{\mu^{2}}b_{4}\left(\frac{\square_{0}}{\mu^2}\right)+\frac{1}{L^{2}}b_{6}\left(\frac{\square_{0}}{\mu^2}\right)+\ldots}\nonumber \\[2mm]
&&+ \ \frac{\frac{1}{2}L^{4}\left(b'_{0}\left(\frac{\square_{2T}-\frac{2}{3}\Lambda}{\mu^2}\right)+b'_{0}\left(\frac{\square_{0}+2\Lambda}{\mu^2}\right)\right)}{\frac{1}{2}L^{4}b_{0}\left(\frac{\square_{0}}{\mu^2}\right)+L^{2}b_{2}\left(\frac{\square_{0}}{\mu^2}\right)+\log\frac{L^{2}}{\mu^{2}}b_{4}\left(\frac{\square_{0}}{\mu^2}\right)+\frac{1}{L^{2}}b_{6}\left(\frac{\square_{0}}{\mu^2}\right)+\ldots}\nonumber \\[2mm]
&&- \ \frac{L^{2}\left(2b'_{2}\left(\frac{\square_{1}+\Lambda}{\mu^2}\right)\right)}{\frac{1}{2}L^{4}b_{0}\left(\frac{\square_{0}}{\mu^2}\right)+L^{2}b_{2}\left(\frac{\square_{0}}{\mu^2}\right)+\log\frac{L^{2}}{\mu^{2}}b_{4}\left(\frac{\square_{0}}{\mu^2}\right)+\frac{1}{L^{2}}b_{6}\left(\frac{\square_{0}}{\mu^2}\right)+\ldots}\nonumber \\[2mm]
&&+ \ \frac{L^{2}\left(b'_{2}\left(\frac{\square_{2T}-\frac{2}{3}\Lambda}{\mu^2}\right)+b'_{2}\left(\frac{\square_{0}+2\Lambda}{\mu^2}\right)\right)}{\frac{1}{2}L^{4}b_{0}\left(\frac{\square_{0}}{\mu^2}\right)+L^{2}b_{2}\left(\frac{\square_{0}}{\mu^2}\right)+\log\frac{L^{2}}{\mu^{2}}b_{4}\left(\frac{\square_{0}}{\mu^2}\right)+\frac{1}{L^{2}}b_{6}\left(\frac{\square_{0}}{\mu^2}\right)+\ldots}\nonumber \\[2mm]
&&- \ \frac{\log\frac{L^{2}}{\mu^{2}}\left(2b'_{4}\left(\frac{\square_{1}+\Lambda}{\mu^2}\right)\right)}{\frac{1}{2}L^{4}b_{0}\left(\frac{\square_{0}}{\mu^2}\right)+L^{2}b_{2}\left(\frac{\square_{0}}{\mu^2}\right)+\log\frac{L^{2}}{\mu^{2}}b_{4}\left(\frac{\square_{0}}{\mu^2}\right)+\frac{1}{L^{2}}b_{6}\left(\frac{\square_{0}}{\mu^2}\right)+\ldots}\nonumber \\[2mm]
&&+ \ \frac{\log\frac{L^{2}}{\mu^{2}}\left(b'_{4}\left(\frac{\square_{2T}-\frac{2}{3}\Lambda}{\mu^2}\right)+b'_{4}\left(\frac{\square_{0}+2\Lambda}{\mu^2}\right)\right)}{\frac{1}{2}L^{4}b_{0}\left(\frac{\square_{0}}{\mu^2}\right)+L^{2}b_{2}\left(\frac{\square_{0}}{\mu^2}\right)+\log\frac{L^{2}}{\mu^{2}}b_{4}\left(\frac{\square_{0}}{\mu^2}\right)+\frac{1}{L^{2}}b_{6}\left(\frac{\square_{0}}{\mu^2}\right)+\ldots}+\ldots\nonumber \\[2mm]
&=& \frac{2+\frac{2}{L^{2}}\Lambda\left(0+\frac{8}{3}-2\times\frac{20}{3}\right)}{1+\frac{2}{L^{2}}b_{2}\left(\frac{\square_{0}}{\mu^2}\right)+\frac{2}{L^{4}}\log\frac{L^{2}}{\mu^{2}}b_{4}\left(\frac{\square_{0}}{\mu^2}\right)+\frac{2}{L^{6}}b_{6}\left(\frac{\square_{0}}{\mu^2}\right)+\ldots}+
\nonumber \\[2mm]
&&+ \ \frac{\frac{2}{L^{4}}\log\frac{L^{2}}{\mu^{2}}\Lambda^{2}\left(-\frac{7}{5}+\frac{479}{135}-2\times\frac{716}{135}\right)+\ldots}{1+\frac{2}{L^{2}}b_{2}\left(\frac{\square_{0}}{\mu^2}\right)+\frac{2}{L^{4}}\log\frac{L^{2}}{\mu^{2}}b_{4}\left(\frac{\square_{0}}{\mu^2}\right)+\frac{2}{L^{6}}b_{6}\left(\frac{\square_{0}}{\mu^2}\right)+\ldots}
\end{eqnarray}
By collecting all terms we get
%
%
%
\begin{eqnarray}
N_{{\rm DOF}} \ = \ \frac{2-\frac{64}{3}\frac{\Lambda}{L^{2}}+\frac{4568}{135}\frac{\Lambda^{2}}{L^{4}}\log\frac{\mu}{L}}{1+\frac{4}{3}\frac{\Lambda}{L^{2}}-\frac{116}{135}\frac{\Lambda^{2}}{L^{4}}\log\frac{\mu}{L}}\, .
\label{EinMSS}
\end{eqnarray}
In deriving (\ref{EinMSS}), we  used the results from~\cite{RJ}, namely for the un-integrated hat kernel coefficients $b_4$'s
\begin{eqnarray}
&&b'_{4}\left(\frac{\square_{2T}-\frac{2}{3}\Lambda}{\mu^2}\right) \ = \ - \frac{7}{5}\Lambda^{2}\, ,
\;\;\;\;
b'_{4}\left(\frac{\square_{2T}-\frac{4}{3}\Lambda}{\mu^2}\right) \ = \ \frac{3}{5}\Lambda^{2}\, ,
\;\;\;\;
b'_{4}\left(\frac{\square_{0}+2\Lambda}{\mu^2}\right) \ = \ \frac{479}{135}\Lambda^{2}\, ,
\nonumber \\[2mm]
&&b'_{4}\left(\frac{\square_{1}+\Lambda}{\mu^2}\right)=\frac{716}{135}\Lambda^{2}\, , \;\;\;\;
b'_{4}\left(\frac{\square_{1}+\frac{1}{3}\Lambda}{\mu^2}\right)=\frac{236}{135}\Lambda^{2}\, ,
\;\;\;
b'_{4}\left(\frac{\square_{0}+\frac{4}{3}\Lambda}{\mu^2}\right)=\frac{269}{135}\Lambda^{2}\, ,
\nonumber \\[2mm]
&&
b_{4}\left(\frac{\square_{0}}{\mu^2}\right)=\frac{29}{135}\Lambda^{2}\, .
\end{eqnarray}
Moreover, for the coefficients $b'_2$ we have from~\cite{RJ}
\begin{eqnarray}
b'_{2}\left(\frac{\square+Y\hat{1}}{\mu^2}\right) &=& \left(\frac{1}{6}R+Y\right){\rm tr}\hat{1} \ = \ \left(\frac{2}{3}\Lambda+Y\right){\rm tr}\hat{1}\, ,
\end{eqnarray}
and similarly 
\begin{eqnarray}
&&b'_{2}\left(\frac{\square_{2T}-\frac{2}{3}\Lambda}{\mu^2}\right) \ = \ 9\left(\frac{2}{3}\Lambda-\frac{2}{3}\Lambda\right) \ = \ 0\, ,
\nonumber \\[2mm]&&
b'_{2}\left(\frac{\square_{2T}-\frac{4}{3}\Lambda}{\mu^2}\right) \ = \ 9\left(\frac{2}{3}\Lambda-\frac{4}{3}\Lambda\right) \ = \ -6\Lambda\, ,
\nonumber \\[2mm]
&&b'_{2}\left(\frac{\square_{0}+2\Lambda}{\mu^2}\right) \ = \ \frac{2}{3}\Lambda+2\Lambda \ = \ \frac{8}{3}\Lambda
\, ,
\nonumber \\[2mm]&&
b'_{2}\left(\frac{\square_{1}+\Lambda}{\mu^2}\right) \ = \ 4\left(\frac{2}{3}\Lambda+\Lambda\right) \ = \ \frac{20}{3}\Lambda
\, ,
\nonumber \\[2mm]
&&
b'_{2}\left(\frac{\square_{1}+\frac{1}{3}\Lambda}{\mu^2}\right)=4\left(\frac{2}{3}\Lambda+\frac{1}{3}\Lambda\right)=4\Lambda
\, ,
\nonumber \\[2mm]&&
b'_{2}\left(\frac{\square_{0}+\frac{4}{3}\Lambda}{\mu^2}\right) \ = \ \frac{2}{3}\Lambda+\frac{4}{3}\Lambda \ = \ 2\Lambda
\, ,
\nonumber \\[2mm]
&&b_{2}\left(\frac{\square_{0}}{\mu^2}\right) \ = \ \frac{2}{3}\Lambda\, ,
\end{eqnarray}
%
%
%
%
Plot of this running for various values of the dimensionless parameter $\lambda$ is depicted in Fig.~\ref{Fig.4}. 
\begin{figure}[h]
\begin{center}
$\mbox{\hspace{4mm}}$\includegraphics[width= 8.0cm]{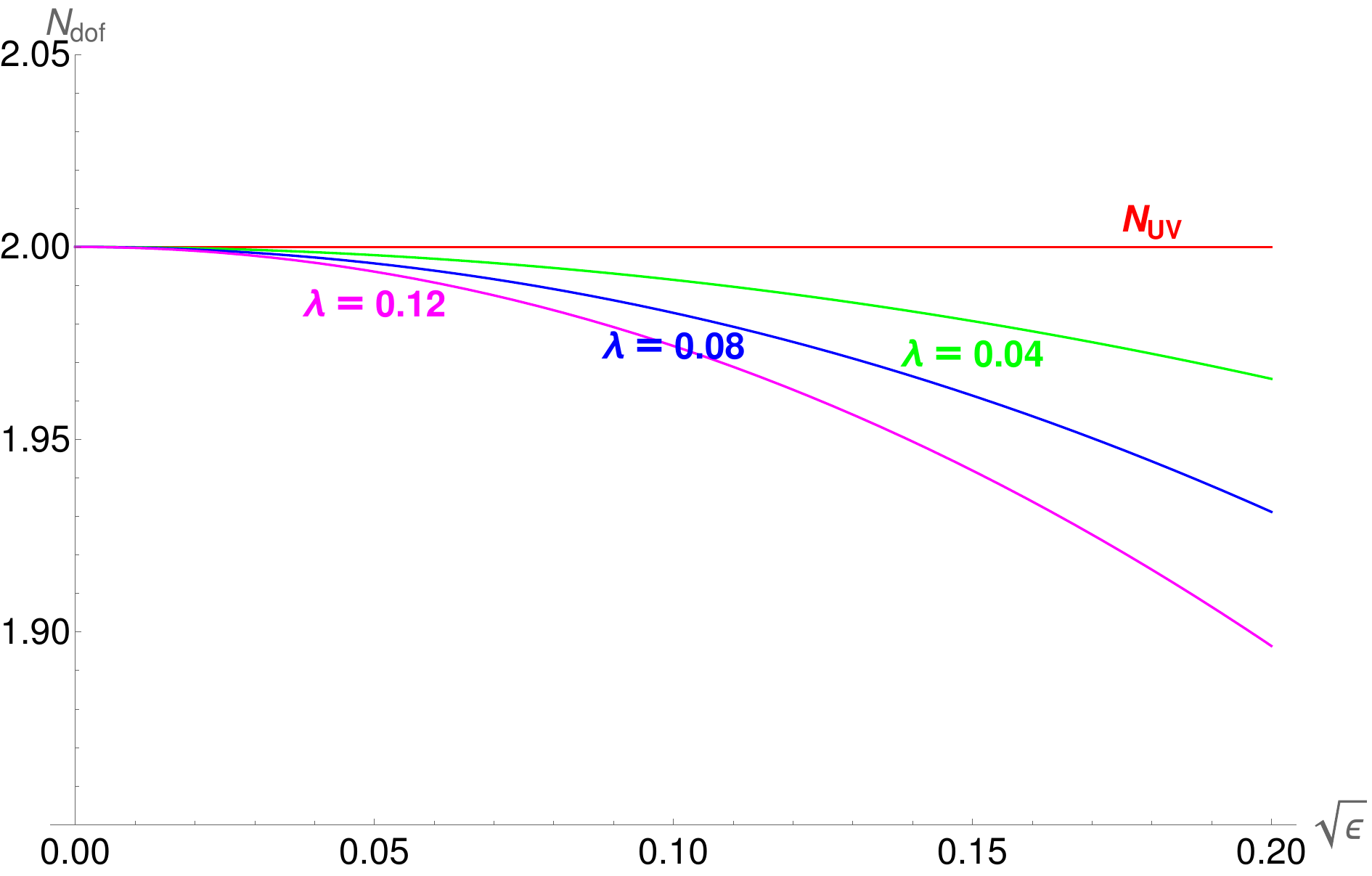}
\vspace{1mm}
\caption{Plot of $N_{\rm DOF}$ as a function of $\sqrt{\epsilon}=\mu / L$ for various values of $\lambda$ in 4-dimensional Einsteinian gravitation on MSS background. In red we show the value of DOF in the UV limit.}
\label{Fig.4}
\end{center}
\end{figure}
\end{widetext}

\subsection{Ricci-flat background}

Properties of a 4-dimensional Ricci-flat background are specified at the beginning of Section~\ref{Sec.IIIB}.
The partition function a Ricci-flat background of EG at the one-loop level reads
\begin{equation}
Z^{2} \ = \ \frac{\det^{2}\left(\frac{\square_{1}}{\mu^2}\right)}{\det\left(\frac{\square_{2T}+2\hat{C}}{\mu^2}\right)\det\left(\frac{\square_{0}}{\mu^2}\right)}\, .
\end{equation}
From this we get the effective number of DOF
\begin{widetext}
\begin{eqnarray}
N_{{\rm DOF}} &=& 1 \ - \ \frac{2\log\det\left(\frac{\square_{1}}{\mu^2}\right)-\log\det\left(\frac{\square_{2T}+2\hat{C}}{\mu^2}\right)}{\log\det\left(\frac{\square_{0}}{\mu^2}\right)} \nonumber \\[3mm] &=&  1-\frac{L^{4}b'_{0}\left(\frac{\square_{1}}{\mu^2}\right)+2L^{2}b'_{2}\left(\frac{\square_{1}}{\mu^2}\right)+2\log\frac{L^{2}}{\mu^{2}}b'_{4}\left(\frac{\square_{1}}{\mu^2}\right)}{\frac{1}{2}L^{4}b_{0}\left(\frac{\square_{0}}{\mu^2}\right)+L^{2}b_{2}\left(\frac{\square_{0}}{\mu^2}\right)+\log\frac{L^{2}}{\mu^{2}}b_{4}\left(\frac{\square_{0}}{\mu^2}\right)}+ \ \frac{\frac{1}{2}L^{4}b'_{0}\left(\frac{\square_{2T}+2\hat{C}}{\mu^2}\right)+L^{2}b'_{2}\left(\frac{\square_{2T}+2\hat{C}}{\mu^2}\right)}{\frac{1}{2}L^{4}b_{0}\left(\frac{\square_{0}}{\mu^2}\right)+L^{2}b_{2}\left(\frac{\square_{0}}{\mu^2}\right)+\log\frac{L^{2}}{\mu^{2}}b_{4}\left(\frac{\square_{0}}{\mu^2}\right)}\nonumber \\[3mm] &&+ \
\frac{\log\frac{L^{2}}{\mu^{2}}b'_{4}\left(\frac{\square_{2T}+2\hat{C}}{\mu^2}\right)}{\frac{1}{2}L^{4}b_{0}\left(\frac{\square_{0}}{\mu^2}\right)+L^{2}b_{2}\left(\frac{\square_{0}}{\mu^2}\right)+\log\frac{L^{2}}{\mu^{2}}b_{4}\left(\frac{\square_{0}}{\mu^2}\right)} \nonumber \\[3mm]
&=& 1 \ - \ \frac{4L^{4}-2\log\frac{\mu^{2}}{L^{2}}b'_{4}\left(\frac{\square_{1}}{\mu^2}\right)-\frac{9}{2}L^{4}+\log\frac{\mu^{2}}{L^{2}}b'_{4}\left(\frac{\square_{2T}+2\hat{C}}{\mu^2}\right)}{\frac{1}{2}L^{4}-\log\frac{\mu^{2}}{L^{2}}b_{4}\left(\frac{\square_{0}}{\mu^2}\right)}\nonumber \\[3mm]
&=& 1 \ - \ \frac{-1-\frac{8}{L^{4}}\log\epsilon b'_{4}\left(\frac{\square_{1}}{\mu^2}\right)+\frac{4}{L^{4}}\log\epsilon b'_{4}\left(\frac{\square_{2}+2\hat{C}}{\mu^2}\right)-\frac{4}{L^{4}}\log\epsilon b_{4}\left(\frac{\square_{0}}{\mu^2}\right)}{1-\frac{4}{L^{4}}\log\epsilon b_{4}\left(\frac{\square_{0}}{\mu^2}\right)}\nonumber \\[3mm]
&=& \frac{2+\frac{8}{L^{4}}\log\epsilon b'_{4}\left(\frac{\square_{1}}{\mu^2}\right)-\frac{4}{L^{4}}\log\epsilon b'_{4}\left(\frac{\square_{2}+2\hat{C}}{\mu^2}\right)}{1-\frac{4}{L^{4}}\log\epsilon b_{4}\left(\frac{\square_{0}}{\mu^2}\right)}\, .
\end{eqnarray}
\end{widetext}
By employing the results from~\cite{RJ} 
\begin{eqnarray}
&&b_{4}\left(\frac{\square_{0}}{\mu^2}\right) \ = \ \frac{1}{180}C^{2}\, ,\;\;\;
b'_{4}\left(\frac{\square_{1}}{\mu^2}\right)\ = \ -\frac{11}{180}C^{2}\, , \nonumber \\[2mm] 
&&b'_{4}\left(\frac{\square_{2}+2\hat{C}}{\mu^2}\right) \ = \ \frac{19}{18}C^{2}\, ,\nonumber \\[2mm]
&& b'_{4}\left(\frac{\square_{2T}+2\hat{C}}{\mu^{2}}\right)=b'_{4}\left(\frac{\square_{2}+2\hat{C}}{\mu^{2}}\right)-b_{4}\left(\frac{\square_{0}}{\mu^{2}}\right)\nonumber \\[2mm]
&&=\frac{19}{18}C^{2}-\frac{1}{180}C^{2}=\frac{189}{180}C^{2}
\end{eqnarray}
we get
\begin{eqnarray}
 \frac{2+\frac{8}{L^{4}}\log\epsilon\left(-\frac{11}{180}C^{2}\right)-\frac{4}{L^{4}}\log\epsilon\frac{19}{18}C^{2}}{1-\frac{4}{L^{4}}\log\epsilon\frac{1}{180}C^{2}}\, .~~~~~
\end{eqnarray}

After using the remark from the appendix \ref{Appendix C} in effectively substituting $C^2$  field with the parameter $\alpha$, we find

\begin{eqnarray}
N_{{\rm DOF}} =  \frac{2+\frac{8}{L^{4}}\epsilon^2\log\epsilon\left(-\frac{11}{180}\alpha\right)-\frac{4}{L^{4}}\epsilon^2\log\epsilon\frac{19}{18}\alpha}{1-\frac{4}{L^{4}}\epsilon^2\log\epsilon\frac{1}{180}\alpha}\, .~~~~~
\end{eqnarray}

Finally, in Einstein's gravity, we find that
\begin{equation}
N_{{\rm DOF}} \ = \ \frac{2-\frac{212}{45}\alpha\left(\frac{\mu}{L}\right)^2\log\frac{\mu}{L}}{1-\frac{1}{45}\alpha\left(\frac{\mu}{L}\right)^2\log\frac{\mu}{L}}\, .\label{EinRic}
\end{equation}
We see that also in the limiting point $\mu=0$ the number of degrees of freedom reaches two as it does for the Einsteinian quantum gravity in the UV.
The plots of this running for various values of the parameter $\alpha$ are depicted in Fig.~\ref{Fig.5}.

\begin{figure}[!ht]
\begin{center}
$\mbox{\hspace{4mm}}$\includegraphics[width= 8.0cm]{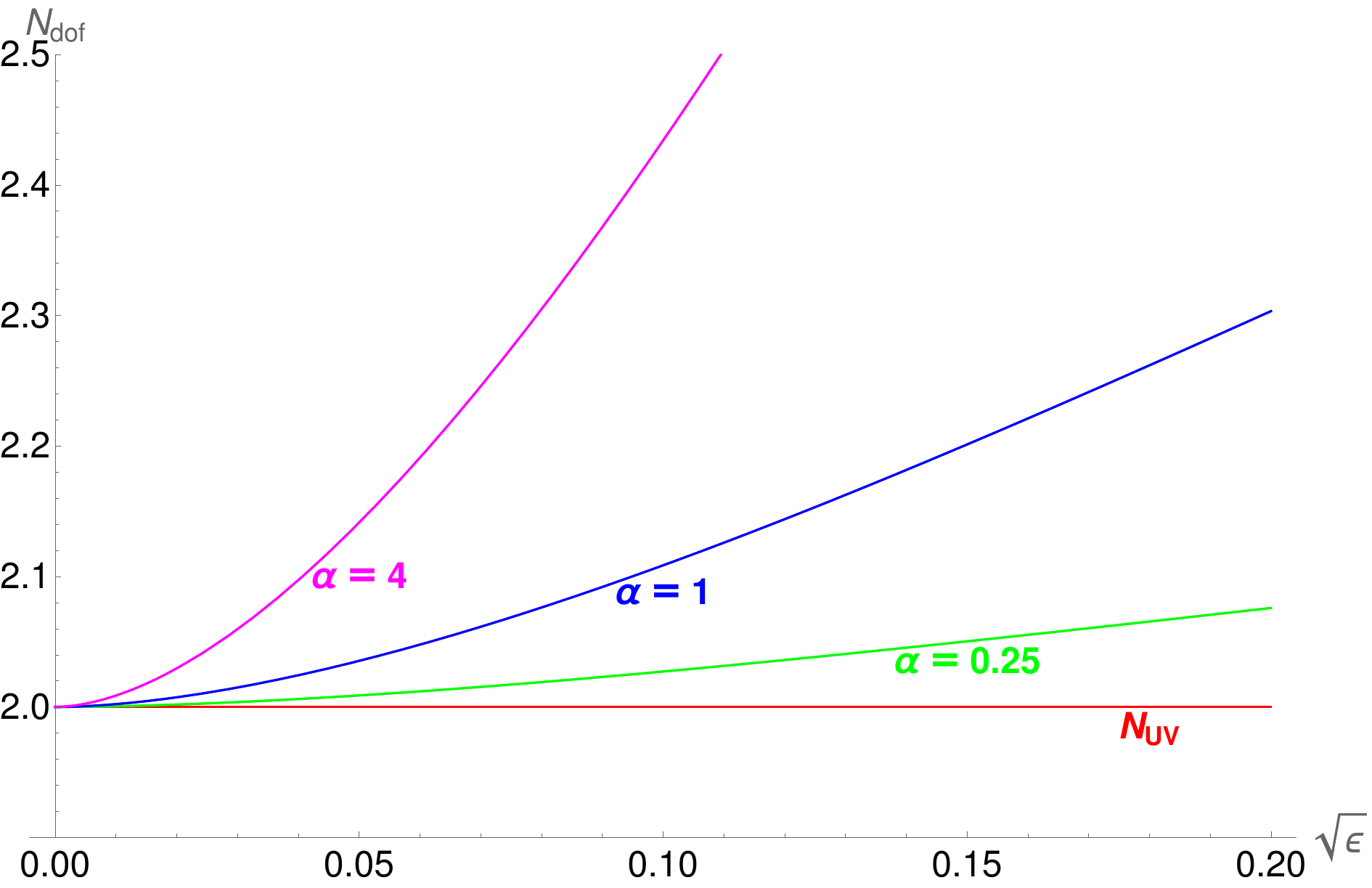}
\vspace{1mm}
\caption{Plot of $N_{\rm DOF}$ as a function of $\sqrt{\epsilon}=\mu / L$ for various values of the dimensionless parameter $\alpha$ in Einsteinian gravitation on Ricci-flat background. In red we show the value of DOF in the UV limit.}
\label{Fig.5}
\end{center}
\end{figure}

%
%
%
%
%

%

Note that both MSS and Ricci-flat backgrounds show a distinct running of DOF from UV to IR.  This once again illustrates the dependence of the FT prescription on the background.

\section{Derivation of Eq.~(\ref{Sec.V.57.nn})  \label{Appendix B}}

Here we derive the relation~(\ref{Sec.V.57.nn}).   We start with the fact that the effective action 
\begin{eqnarray}
\Gamma(g^{(0)}) &\equiv& \Gamma(J=0,g^{(0)}) \ = \ -\log Z(g^{(0)}) \nonumber \\[2mm] 
&=&  \frac{1}{2} {\rm{Tr}} \log \tilde{\Delta}\, ,
\end{eqnarray}
where $g_{\mu\nu}^{(0)}$ is a background metric, $J_{\mu\nu}$ is a source term and 
\begin{eqnarray}
\tilde{\Delta} \ = \ \frac{\Delta}{\mu^4}\, ,
\end{eqnarray}
is the (quartic) kernel operator from the classical action. 
In order to make sense of the trace term, we use the $\zeta$-function regularization. To this end we write
\begin{eqnarray}
\frac{1}{2} {\rm{Tr}} \log \tilde{\Delta} \ = \ \frac{1}{2} \sum_n \log \tilde{\lambda}_n\, ,
\end{eqnarray}
where $\tilde{\lambda}_n$ are the dimensionless eigenvalues. We define a $\zeta$-function of the kernel $\Delta$ as
\begin{eqnarray}
\zeta_{\Delta}(s) \ = \ \sum_{n} \tilde{\lambda}^{-s}_n \, .
\end{eqnarray}
By using the replica-trick identity
\begin{eqnarray}
\log x \ = \ - \left.\frac{d}{d s} x^{-s} \right|_{s = 0}\, ,
\end{eqnarray}
we have
\begin{eqnarray}
\frac{1}{2} {\rm{Tr}} \log \tilde{\Delta} \ = \ \left.-\frac{1}{2} 
\zeta_{\Delta}(s)\right|_{s=0}\, .
\end{eqnarray}
Let us now consider $\Gamma(\Omega^2 g^{(0)})$ where $\Omega^2$ is a constant.   For the latter case we can write
\begin{eqnarray}
\Gamma({\Omega^2 g^{(0)}}) &=& \frac{1}{2} \sum_n \log\left(\frac{\Omega^{-4} \lambda_n}{\mu^4} \right)\nonumber \\[2mm]
&=& \frac{1}{2} \sum_n \log \left(\frac{ \lambda_n}{\mu^4} \right) \ + \ \frac{1}{2} \log \left(\Omega^{-4} \right) {\rm{Tr} \ \!\hat{\mathbb{1}}} \nonumber \\[2mm]
&=& \Gamma(g^{(0)}) \ - \ 2 \log \Omega \ \! \zeta_{\Delta} (0)\, ,
\label{B7.cc}
\end{eqnarray}
where $\Omega^{-2}$ on the first line appears because $\Omega^2$ multiplies the covariant metric tensor ${g^{(0)}}_{\!\!\mu\nu}$. The term $\zeta_{\Delta}(0)$ is finite when analytically continued from $\rm{Re}(s) > 2$. Relation~(\ref{B7.cc}) is the sought result. 

\section{Derivation of Eq.~(\ref{form26}) \label{Appendix C}}

We use the following Euclidean Schwarzschild metric
\begin{eqnarray}
ds^{2}&=&\left(1+\frac{2M}{r}\right)d\tau^{2} \ + \ \frac{1}{1+\frac{2M}{r}}dr^{2}\nonumber \\[2mm]
&&+ \ r^{2}d\theta^{2} 
\ + \ r^{2}\sin^{2}\theta d\phi^{2}\, ,
\label{metricESchw}
\end{eqnarray}
for the unbounded and non-compact 4-dimensional Euclidean manifold. This is still a vacuum solution for conformal gravitation (as well as for Einsteinian gravitation), but the flip of the sign in front of the mass parameter $M$ suggests it is a rather anti-gravitational Euclidean Schwarzschild black hole solution. This 4-dimensional space has to be considered with the following range of the radial coordinate: $r\geqslant0$.
The origin $r=0$ is a singular point. The Ricci tensor vanishes everywhere $R_{\mu\nu}=0$
and this implies that also $R=0$ everywhere. We remark that the purported location of the ``horizon'' at $r=2M$ is not problematic here, the Euclidean space manifold is here smooth and the signature of the metric does not change sign (in accordance with Sylvester's law of inertia for the metric). Hence, this manifold is only singular at the point $r=0$ even if one uses Schwarzschild-like type of coordinates. There is no any coordinate singularity in this system for $r>0$, but of course, one could use better suited Kruskal--Szekeres type of coordinates to display the global structure of the manifold. In the following we will consider invariants where the results of the computation are not obviously dependent on the particular coordinate system chosen. We use the metric (\ref{metricESchw}) as a particular example of the Ricci-flat background to compute the RG flow and the resulting change in the number of DOF with the energy scale, and we do not assign any physical or geometrical interpretation to it.

The Kretschmann invariant reads
\begin{equation}
K \ = \ C^{2} \ = \ \frac{48M^{2}}{r^{6}}\, ,
\end{equation}
and it is singular in the limit $r\to0$.

The proper volume spacetime element is
\begin{equation}
\sqrt{|g|} \ = \ \sqrt{\det g} \ = \ r^{2}\sin\theta\, .
\end{equation}
The general ratio, which appears with the computation using  HK coefficients reads,
\begin{equation}
\mathsf{R} \ = \ \frac{\int\! d^{4}x\sqrt{|g|}\left(A_{n}+B_{n}C^{2}\right)}{\int\! d^{4}x\sqrt{|g|}\left(A_{d}+B_{d}C^{2}\right)},
\label{ratio1}
\end{equation}
where $A_n$ and $B_n$ denote some constant space-independent coefficients of the heat kernel expansion in the numerator of the ratio $\mathsf{R}$ (and the index ``$d\,$'' for the respective coefficients in the denominator). The whole space dependence is contained only in the position-dependent scalar field $C^2=C^2(x)$  and in the densities $\sqrt{|g|}$.

In the formula (\ref{ratio1}) we simplify the constant $\tau$-integration, because here nothing
depends on the $\tau$ coordinate, and the ratio is expressed 
as
\begin{eqnarray}
\mathsf{R}&=&\frac{4\pi\int\! drr^{2}\left(A_{n}+B_{n}\frac{48M^{2}}{r^{6}}\right)}{4\pi\int\! drr^{2}\left(A_{d}+B_{d}\frac{48M^{2}}{r^{6}}\right)}\nonumber \\[2mm]
&=&\frac{\int\! drr^{2}\left(A_{n}+B_{n}\frac{48M^{2}}{r^{6}}\right)}{\int\! drr^{2}\left(A_{d}+B_{d}\frac{48M^{2}}{r^{6}}\right)}\, .
\end{eqnarray}
When the radial integral is performed in the limits $r_{-}$ and $+\infty$,
then the ratio $\mathsf{R}$ gives $\frac{A_{n}}{A_{d}}$. In turn
if in the limits $0$ to $r_{+}$, then the ratio $\mathsf{R}$ gives
$\frac{B_{n}}{B_{d}}$. Finally, if in the limits $r_{-}$ and $r_{+}>r_{-}>0$
(both finite), then
\begin{widetext}
\begin{equation}
\mathsf{R}\ = \ \frac{\left[A_{n}\frac{r^{3}}{3}-16B_{n}\frac{M^{2}}{r^{3}}\right]_{r_{-}}^{r_{+}}}{\left[A_{d}\frac{r^{3}}{3}-16B_{d}\frac{M^{2}}{r^{3}}\right]_{r_{-}}^{r_{+}}}\ = \ \frac{A_{n}\left(r_{+}^{3}-r_{-}^{3}\right)-48B_{n}M^{2}\left(\frac{1}{r_{+}^{3}}-\frac{1}{r_{-}^{3}}\right)}{A_{d}\left(r_{+}^{3}-r_{-}^{3}\right)-48B_{d}M^{2}\left(\frac{1}{r_{+}^{3}}-\frac{1}{r_{-}^{3}}\right)} \ = \ \frac{r_{+}^{3}r_{-}^{3}A_{n}+48B_{n}M^{2}}{r_{+}^{3}r_{-}^{3}A_{d}+48B_{d}M^{2}}\, .\end{equation}
\end{widetext}

The constant term ($A_n$ or $A_d$) causes the problem with the large $r$ integration
(for the infinite spacetime) -- IR problem, while the Kretschmann term
is problematic in the small $r$ regime for the integration (near
the singularity) -- UV problem. Since we have both at the same time,  it is difficult to regulate them simultaneously.

For explicit integration, we could choose, for example
\begin{equation}
r_{-} \ = \ 2M\quad{\rm and}\quad r_{+} \ = \ 3M\, , \end{equation}
and then the ratio $\mathsf{R}$ simplifies to
\begin{equation}
\mathsf{R}\ = \ \frac{9M^{6}A_{n}+2B_{n}M^{2}}{9M^{6}A_{d}+2B_{d}M^{2}}\ = \ \frac{9M^{4}A_{n}+2B_{n}}{9M^{4}A_{d}+2B_{d}}\, .\end{equation}
This can be put back in the form
\begin{equation}
\mathsf{R} \ = \ \frac{A_{n}+\frac{2}{9}\frac{1}{M^{4}}B_{n}}{A_{d}+\frac{2}{9}\frac{1}{M^{4}}B_{d}}\, ,\end{equation}
which effectively looks like the ratio
\begin{equation}
\frac{A_{n}+B_{n}C^{2}}{A_{d}+B_{d}C^{2}}\, ,\end{equation}
with the identification of $C^{2}$ with $\frac{2}{9}\frac{1}{M^{4}}$.

By defining 
\begin{equation}
\tilde{\alpha} \ = \ \frac{48M^{2}}{r_{+}^{3}r_{-}^{3}}\, ,\end{equation}
we can write that the ratio takes the following dependence on the
parameter $\tilde{\alpha}$
\begin{equation}
\mathsf{R} \ = \ \frac{A_{n}+\tilde{\alpha} B_{n}}{A_{d}+\tilde{\alpha} B_{d}}\, ,\end{equation}
in the general case. Therefore, it is sufficient to plot the ratio $\mathsf{R}$
for different values of the parameter $\tilde{\alpha}$ and in this way we can forget about the problem of the dependence of the volume spacetime integration in the case of Ricci-flat backgrounds. Effectively we can just plot
the ratios of densities with arbitrary values of the constant parameter
$\tilde{\alpha}$.

Eventually, since the parameter $\tilde{\alpha}$ is dimensionful we can redefine it with the help of the relation $\tilde{\alpha}=\alpha\mu^4$, where now the parameter $\alpha$ is properly dimensionless and it does not depend on the space point.

\end{document}